\documentclass[twocolumn]{aastex6311}
\usepackage{threeparttable}
\usepackage{multirow}
\usepackage{times}
\usepackage{enumitem}
\usepackage{natbib}
\usepackage{url} 
\usepackage{amsmath,bm}
\usepackage{multirow}
\usepackage{xcolor}
\definecolor{hyblue}{RGB}{0,0,200} 
\definecolor{hyred}{RGB}{200,0,0}

\shorttitle{Stellar Parameters and Abundances for S-PLUS DR4 Stars}
\shortauthors{Yang Huang et al.}

\begin{document}

\title{S-PLUS: Beyond Spectroscopy IV. Stellar Parameters and Elemental-abundance Ratios for Six Million Stars from DR4 and First Results for the Magellanic Clouds}

\author[0000-0003-3250-2876]{Yang Huang}
\affiliation{School of Astronomy and Space Science, University of Chinese Academy of Sciences, Beijing 100049, China}
\affiliation{National Astronomical Observatories, Chinese Academy of Sciences, Beijing 100101, China}

\author[0000-0003-4573-6233]{Timothy C. Beers}
\affiliation{Department of Physics and Astronomy, University of Notre Dame, Notre Dame, IN 46556, USA}
\affiliation{Joint Institute for Nuclear Astrophysics -- Center for the Evolution of the Elements (JINA-CEE), USA}

\author[0000-0001-8424-1079]{Kai Xiao}
\affiliation{School of Astronomy and Space Science, University of Chinese Academy of Sciences, Beijing 100049, China}

\author[0000-0002-5267-9065]{C. Mendes de Oliveira}
\affiliation{Departamento de Astronomia, Instituto de Astronomia, Geof{\'i}sica e Ci{\^e}ncias Atmosf{\'e}ricas, Universidade de S{\~a}o Paulo,\\Rua do Mat{\~a}o 1226, Cidade Universit{\'a}ria, 05508-900 S{\~a}o Paulo, SP, Brazil}

\author[0000-0002-8048-8717]{Felipe Almeida-Fernandes}
\affiliation{Observatório do Valongo, Ladeira Pedro Antonio, 43, 20080-090, Rio de Janeiro, Brazil}
\affiliation{Instituto Nacional de Pesquisas Espaciais, Av. dos Astronautas 1758, Jardim da Granja,12227-010 São José dos Campos, SP, Brazil}

\author[0009-0003-6609-1582]{G.B. Oliveira Schwarz}
\affiliation{Departamento de Astronomia, Instituto de Astronomia, Geof{\'i}sica e Ci{\^e}ncias Atmosf{\'e}ricas, Universidade de S{\~a}o Paulo,\\Rua do Mat{\~a}o 1226, Cidade Universit{\'a}ria, 05508-900 S{\~a}o Paulo, SP, Brazil}

\author[0000-0001-5297-4518]{Young Sun Lee}
\affiliation{Department of Astronomy and Space Science, Chungnam National University, Daejeon 34134, Republic of Korea}

\author[0000-0002-2453-0853]{Jihye Hong}
\affiliation{Department of Physics and Astronomy, University of Notre Dame, Notre Dame, IN 46556, USA}
\affiliation{Joint Institute for Nuclear Astrophysics -- Center for the Evolution of the Elements (JINA-CEE), USA}

\author[0009-0008-2988-2680]{Huiling Chen}
\affiliation{Department of Astronomy, School of Physics, Peking University, Beijing 100871, China}
\affiliation{Kavli Institute for Astronomy and Astrophysics, Peking University, Beijing 100871, China}

\author[0000-0002-7727-1699]{Huawei Zhang}
\affiliation{Department of Astronomy, School of Physics, Peking University, Beijing 100871, China}
\affiliation{Kavli Institute for Astronomy and Astrophysics, Peking University, Beijing 100871, China}

\author[0000-0002-9269-8287]{Guilherme~Limberg}
\affiliation{Kavli Institute for Cosmological Physics, University of Chicago, 5640 S Ellis Avenue, Chicago, IL 60637, USA}

\author[0009-0008-9042-4478]{Maiara S. Carvalho}
\affiliation{Departamento de Astronomia, Instituto de Astronomia, Geof{\'i}sica e Ci{\^e}ncias Atmosf{\'e}ricas, Universidade de S{\~a}o Paulo,\\Rua do Mat{\~a}o 1226, Cidade Universit{\'a}ria, 05508-900 S{\~a}o Paulo, SP, Brazil}

\author[0000-0003-3537-4849]{P. K. Humire}
\affiliation{Departamento de Astronomia, Instituto de Astronomia, Geof{\'i}sica e Ci{\^e}ncias Atmosf{\'e}ricas, Universidade de S{\~a}o Paulo,\\Rua do Mat{\~a}o 1226, Cidade Universit{\'a}ria, 05508-900 S{\~a}o Paulo, SP, Brazil}

\author[0009-0007-1625-8937]{Andr\'e Luiz Figueiredo}
\affiliation{Departamento de Astronomia, Instituto de Astronomia, Geof{\'i}sica e Ci{\^e}ncias Atmosf{\'e}ricas, Universidade de S{\~a}o Paulo,\\Rua do Mat{\~a}o 1226, Cidade Universit{\'a}ria, 05508-900 S{\~a}o Paulo, SP, Brazil}

\author[0000-0003-4254-7111]{Bruno Dias}
\affiliation{Instituto de Astrof\'{\i}sica, Departamento de F\'{\i}sica y Astronom\'{\i}a, 
Facultad de Ciencias Exactas, Universidad Andrés Bello, Santiago, Chile}

\author[0000-0002-5045-9675]{Alvaro Alvarez-Candal}
\affiliation{Instituto de Astrofísica de Andalucía, CSIC, Apt 3004, 18080 Granada, Spain}
\affiliation{Instituto de Física Aplicada a las Ciencias y las Tecnologías, Universidad de Alicante, San Vicent del Raspeig, 03080 Alicante, Spain}

\author[0000-0002-7865-3971]{Marcos Fonseca-Faria}
\affiliation{Laboratório Nacional de Astrofísica, Rua dos Estados Unidos, 154, CEP 37504-364, Itajubá MG, Brazil}

\author[0000-0002-2484-7551]{A. Kanaan}
\affiliation{Departamento de Física, Universidade Federal de Santa Catarina, Florianópolis, 88040-900, SC, Brazil}

\author[0000-0002-0138-1365]{T. Ribeiro}
\affiliation{Rubin Observatory Project Office, 950 N. Cherry Ave, Tucson 85719, USA}

\author[0000-0002-4064-7234]{W. Schoenell}
\affiliation{The Observatories of the Carnegie Institution for Science, 813 Santa Barbara St, Pasadena, CA 91101, USA}

\author[0000-0001-7479-5756]{Silvia Rossi}
\affiliation{Departamento de Astronomia, Instituto de Astronomia, Geof{\'i}sica e Ci{\^e}ncias Atmosf{\'e}ricas, Universidade de S{\~a}o Paulo,\\Rua do Mat{\~a}o 1226, Cidade Universit{\'a}ria, 05508-900 S{\~a}o Paulo, SP, Brazil}

\begin{abstract}
We combine narrow/medium-band filter photometry from the Southern Photometric Local Universe Survey (S-PLUS) DR4 with ultra broad-band filter photometry from \textit{Gaia} EDR3 to derive fundamental stellar parameters  ($T_{\rm eff}$, 
$\log g$, [Fe/H], ages) and elemental-abundance ratios ([C/Fe] and [$\alpha$/Fe]) for 5.4 million stars in the Galaxy (4.9 million dwarfs and 0.5 million giants), as well as for over 0.7 million red giant stars in the Large and Small Magellanic Clouds (LMC and SMC). 
The precisions of the abundance estimates range from 0.05--0.10\,dex for metallicity in the relatively metal-rich range
([Fe/H]~$> -1.0$) to 0.10--0.30\,dex in the metal-poor regime ([Fe/H]~$<-1.0$), 0.10--0.20\,dex for [C/Fe], and 0.05\,dex for [$\alpha$/Fe]. The stellar parameters for LMC and SMC member stars are somewhat less precise than those from the S-PLUS main survey, primarily because of the effect of high reddening.
The use of both metallicity- and carbon-sensitive filters provides unbiased measurements of both [Fe/H] and [C/Fe], of particular importance for very low-metallicity ([Fe/H] $< -2.0$) stars, where carbon enhancement can lead to systematically high estimates of [Fe/H] when only a single metallicity-sensitive filter is employed.  Furthermore, multiple narrow-band filters enable metallicity estimates down to [Fe/H] $\sim -4.0$ with an accuracy of around 0.3\,dex, exceeding the precision typically achieved by low/medium-resolution spectroscopy. This extensive photometric dataset, combined with the other three datasets in this series, will serve as a valuable legacy resource for Milky Way and Magellanic Clouds studies.
\end{abstract}

\section{Introduction}
Accurate and precise measurement of multi-dimensional parameters for large numbers of stars in the Milky Way (MW) is fundamental to unraveling the assembly and chemical-evolution history of our Galaxy. 
Over the past few decades, this field has undergone two major revolutions. The first revolution was driven by large-scale Galactic spectroscopic surveys such as RAVE \citep{Steinmetz2006}, SDSS/SEGUE \citep{segue1, segue2}, LAMOST \citep{legue,2012RAA....12..723Z}, and APOGEE \citep{apogee}. These surveys have obtained spectroscopic observations for tens of millions of stars, sampling roughly 0.01\% of the stellar populations of the MW. By providing precise radial velocities, atmospheric parameters, and elemental abundances, these datasets have significantly advanced our understanding of the structure, formation, and chemical evolution of the Galaxy.  

The second revolution is being led by ultraviolet and narrow/medium-band photometric surveys that bypass traditional spectroscopy, such as the Pristine survey \citep{Starkenburg2017}, SkyMapper \citep{SMSSDR1, SMSSDR2}, J/S-PLUS \citep{jplus, splus}, SAGES \citep{SAGES}, MAGIC survey \citep{2025arXiv250403593B,2025ApJ...991..101P} and $Gaia$ XP \citep{2023A&A...674A...2D, 2025arXiv250701939Z}. These surveys enable the derivation of fundamental stellar parameters\footnote{The combination of medium/narrow-band filters in the near-ultraviolet with traditional broad-band optical filters provides sensitivity to stellar surface gravity (through the Balmer jump at around \(3646\,\text{\AA}\)) and metallicity (via the line blanketing effect, particularly prominent at wavelengths shorter than \(4300\,\text{\AA}\)).} for hundreds of millions of stars -- an order of magnitude increase in sampling compared to LAMOST -- increasing the fraction of the stellar populations of the MW with such information to approximately 0.1\%.  \citep[e.g.,][]{PaperI, PaperII, PaperIII, Andrae+23}. Furthermore, present and upcoming next-generation surveys, including J-PAS \citep{jpas} and the Chinese Space Station Telescope \citep[CSST;][]{2011SSPMA..41.1441Z}, are expected to push this even further, potentially mapping billions or even tens of billions of stars.  
When combined with the all-sky astrometric data from \textit{Gaia}, which already provides positions, parallaxes, and proper motions for more than a billion stars (about 0.5-1\% of the Galaxy's stellar populations; \citealt{GEDR3}), these comprehensive datasets promise an unprecedented view of the formation and chemical-evolution history of the MW.  

The ``Beyond Spectroscopy" series has already contributed significantly to this effort.  
In the first three papers of this series \citep[][hereafter Papers~I, II, and III]{PaperI, PaperII, PaperIII}, we derived stellar parameters, particularly metallicity, for nearly 50 million stars across roughly \(3\pi\) steradians of the sky (about 75\% of the sky), by combining $uv$ narrow-band photometry from the SAGES DR1 catalog in the Northern sky \citep{SAGES} and the SkyMapper catalog in the Southern sky \citep[SMSS;][]{SMSSDR1, SMSSDR2} with \textit{Gaia} EDR3 broad-band photometry and astrometry \citep{GEDR3}.  
Most recently, using the fourth release of SMSS, \citet{2025RNAAS...9...74H} extended this work to derive stellar parameters for more than 50 million stars in the Southern sky, doubling the number of stars compared to \citet{PaperI}.  
Additionally, using data from J-PLUS DR3, which employs seven specially designed narrow/medium-band and five SDSS-like broad-band filters, Paper~III derived not only stellar atmospheric parameters but also elemental-abundance ratios (e.g., [C/Fe], [Mg/Fe], [$\alpha$/Fe]) for more than five million stars using a Kernel Principal Component Analysis (KPCA)-based machine learning approach. Moreover, the systematic errors in photometric metallicities caused by molecular carbon bands in SMSS and SAGES (which previously relied on just two blue narrow/medium-band filters) have now been mitigated, thanks to the introduction of a carbon-sensitive filter.  

In this work, we present estimates of stellar atmospheric parameters, elemental-abundance ratios, and physical parameters for more than 6 million stars in the Southern sky, including approximately 0.7 million stars in the Magellanic Clouds, using data from the fourth release of S-PLUS \citep{2024A&A...689A.249H} and $Gaia$ EDR3 \citep{GEDR3}. We note that recent studies have also derived stellar atmospheric parameters from S-PLUS data: \cite{2021ApJ...912..147W} analyzed 0.7 million stars from S-PLUS DR2, including [C/Fe], using machine learning techniques based on artificial neural networks, while \cite{2025A&A...693A.306F} estimated atmospheric parameters and several elemental abundance ratios for 5 million stars from S-PLUS DR4 using neural networks and random forest techniques. However, both studies are limited in that they cannot reliably determine metallicities below [Fe/H] $\sim -2.5$. To our knowledge, the present work is the first to derive comprehensive stellar parameters and elemental abundance ratios for such a large sample of stars from S-PLUS. S-PLUS uses 12 filters that are essentially the same as those in J-PLUS \citep{2012SPIE.8450E..3SM}, consisting of seven narrow/medium-band filters (FWHM $100$–$400\,\text{\AA}$; $J0378$, $J0395$, $J0410$, $J0430$, $J0515$, $J0660$, and $J0861$) and five broad-band filters, including the Javalambre $u$ (a revised SDSS $u$) and four SDSS-like filters ($griz$). The seven narrow/medium-band filters were specifically designed to trace prominent stellar absorption features, including the Ca\,\textsc{ii} H+K lines, the molecular CH $G$-band, H$\delta$, the Mg\,\textsc{i}b triplet, H$\alpha$, and the Ca\,\textsc{i} triplet.

Similarly to Paper~III. we first establish relations between stellar colors from S-PLUS and Gaia~EDR3 using the KPCA technique, with stellar labels from the carefully calibrated testing dataset in Paper~III.
The structure of this paper is as follows. Section~2 introduces the main data used in this work, including the construction of the training sets. Section~3 presents the estimates of stellar parameters and elemental abundance ratios, together with the corresponding performance assessments. Section~4 describes the final sample and provides the derived effective temperatures, distances, ages, and surface gravities. Section~5 concludes with a summary and a discussion of future prospects.

\vskip 1cm
\section{Data}

\subsection{S-PLUS DR4}
S-PLUS DR4 includes 1,629 pointings that cover more than 3,000 square degrees of the Southern sky. The observations were conducted using the T80-South telescope located at the Cerro Tololo Inter-American Observatory (CTIO) in Chile. The survey reaches a limiting magnitude (at signal-to-noise ratio = 3) of approximately 19.5 in the blue filters and 21.5 in the red filters. The released catalog provides both aperture photometry (PStotal\footnote{Here, PStotal refers to the 3$\arcsec$ aperture magnitude measured in dual mode, with an aperture correction applied so that it represents the total magnitude of a point source.}) and point-spread function (PSF) photometry, encompassing the main survey area, the Magellanic Clouds (MCs), and the disk survey, with PSF photometry available across all regions and PStotal photometry available for the main and MCs fields \citep{2024A&A...689A.249H}.
The official photometric calibration of DR4 is described in \citet{2022MNRAS.511.4590A}, based on a comprehensive approach that combines multi-band reference catalogs and synthetic stellar models. The photometric zero-points were later refined by \citet{2024ApJS..271...41X} using the improved $Gaia$ XP synthetic photometry (XPSP) method, reducing the typical zero-point uncertainties from 10–20 milli-magnitudes (mmags) (as obtained in the original pipeline) to 1–6 mmag. Unless otherwise noted, photometry with the revised zero-points of \citet{2024ApJS..271...41X} is adopted throughout this study.
In this study, we focus on the main survey based on PStotal photometry (dual mode), as its 90\% completeness limit is 1–2 magnitudes fainter than that of PSF photometry (see Figure 10 of \citealt{2024A&A...689A.249H}).
The disk survey is excluded because its heavy reddening is difficult to correct.
For the MCs, we combine PStotal and PSF photometry to maximize the number of reliably detected sources.

\subsection{$Gaia$ EDR3} \label{sec:GEDR3}
This study also includes $Gaia$ EDR3 ultra-wideband photometry ($G$, $G_{\rm BP}$, and $G_{\rm RP}$). The $Gaia$ EDR3 catalog \citep{GEDR3} provides, in addition to photometry in these bands, high-precision astrometric measurements, i.e, parallaxes and proper motions, for approximately 1.5 billion sources as faint as $G \sim 21$. However, its completeness becomes increasingly complex toward the faint end \citep[see][]{2021A&A...649A...3R}. The $G$-band maintains remarkable photometric precision, with uncertainties of only a few mmag even at $G = 20$. In comparison, the uncertainties for $G_{\rm BP}$ and $G_{\rm RP}$ are typically around 10 mmag at $G = 17$, gradually increasing to no more than 100 mmag at $G = 20$.

By cross-matching S-PLUS DR4 with \textit{Gaia} EDR3 and applying the selection criteria 
\texttt{CLASS = 1}\footnote{\texttt{CLASS = 1} indicates that the source is classified as a star \citep{2021MNRAS.507.5847N}.}, \texttt{S2N\_DET\_AUTO $\geq$ 3}, detection in all 12 filters, and Galactic latitude 
$|b| \geq 10^{\circ}$, we retain more than $8$ million stars from the S-PLUS DR4 main survey. 
For the MC regions, additional membership selection based on proper motion and parallax 
(see Appendix) is applied on top of the above cuts, yielding nearly $2$ million candidate 
member stars. 
For extinction correction in the main survey region, we adopted the 
\citet[][hereafter SFD\footnote{The reported $14\%$ systematic overestimation in extinction 
has been corrected \citep{2011ApJ...737..103S, 2013MNRAS.430.2188Y}.}]{SFD98} map, which is appropriate for these high-latitude fields 
($|b| \geq 10^{\circ}$). For the MC regions, we adopt the extinction map of \citet{2021ApJS..252...23S}. 
The reddening coefficients for the S-PLUS pass-bands 
($k_{\chi} = A_{\chi}/E(B-V)$; see Table~1) are derived from the extinction curve of 
\citet{2016ApJ...821...78S} with $R_{V} = 3.1$, while those for the \textit{Gaia} EDR3  pass-bands are taken from \citet{2021ApJ...907...68H}.

\begin{figure}
\begin{center}
\includegraphics[scale=0.32,angle=0]{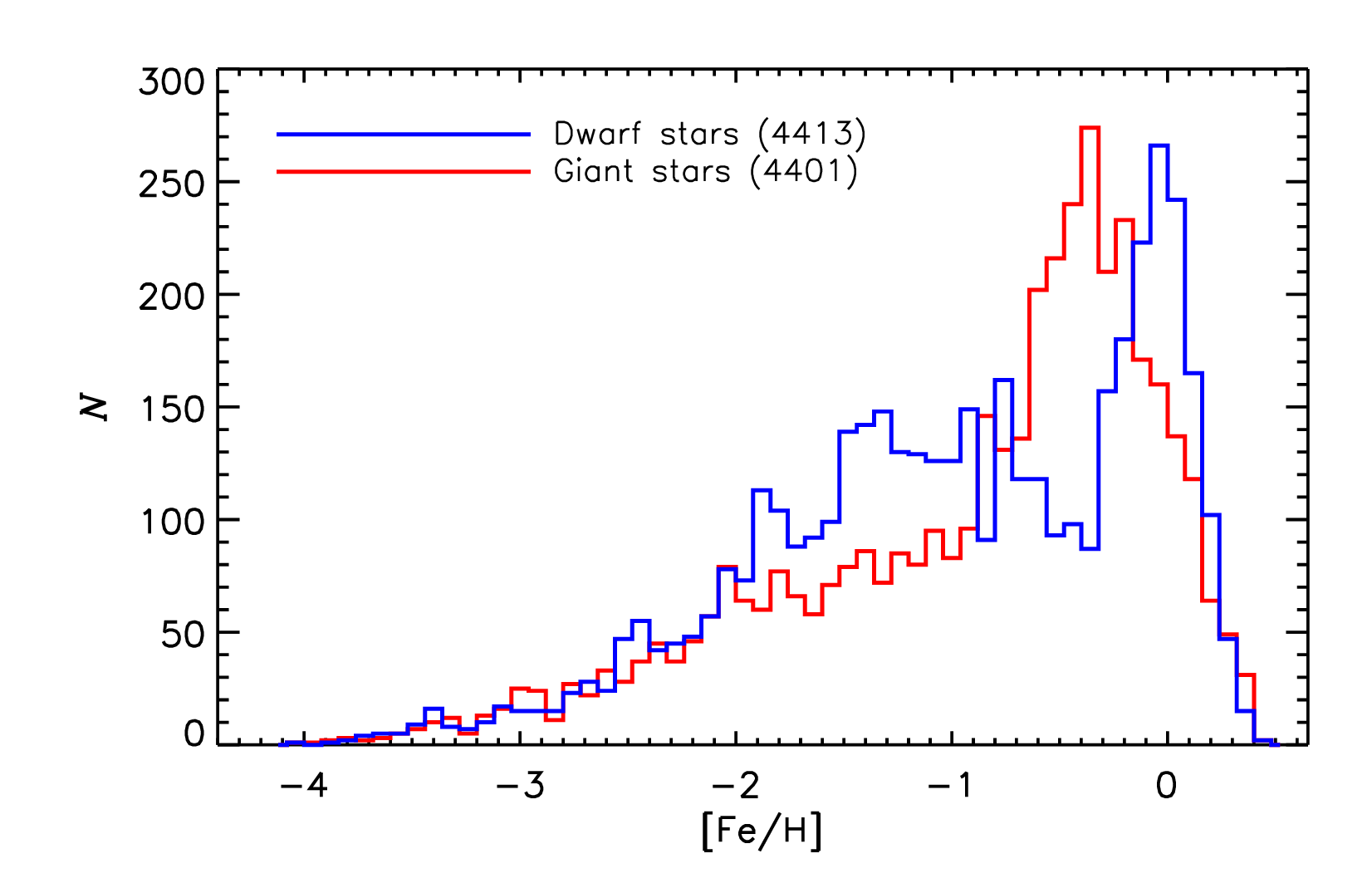}
\caption{Histograms of the training sets in metallicity ([Fe/H]) for dwarfs (blue) and giants (red).}
\label{fig:training_feh}
\end{center}
\end{figure} 

\begin{figure*}
\begin{center}
\includegraphics[scale=0.38,angle=0]{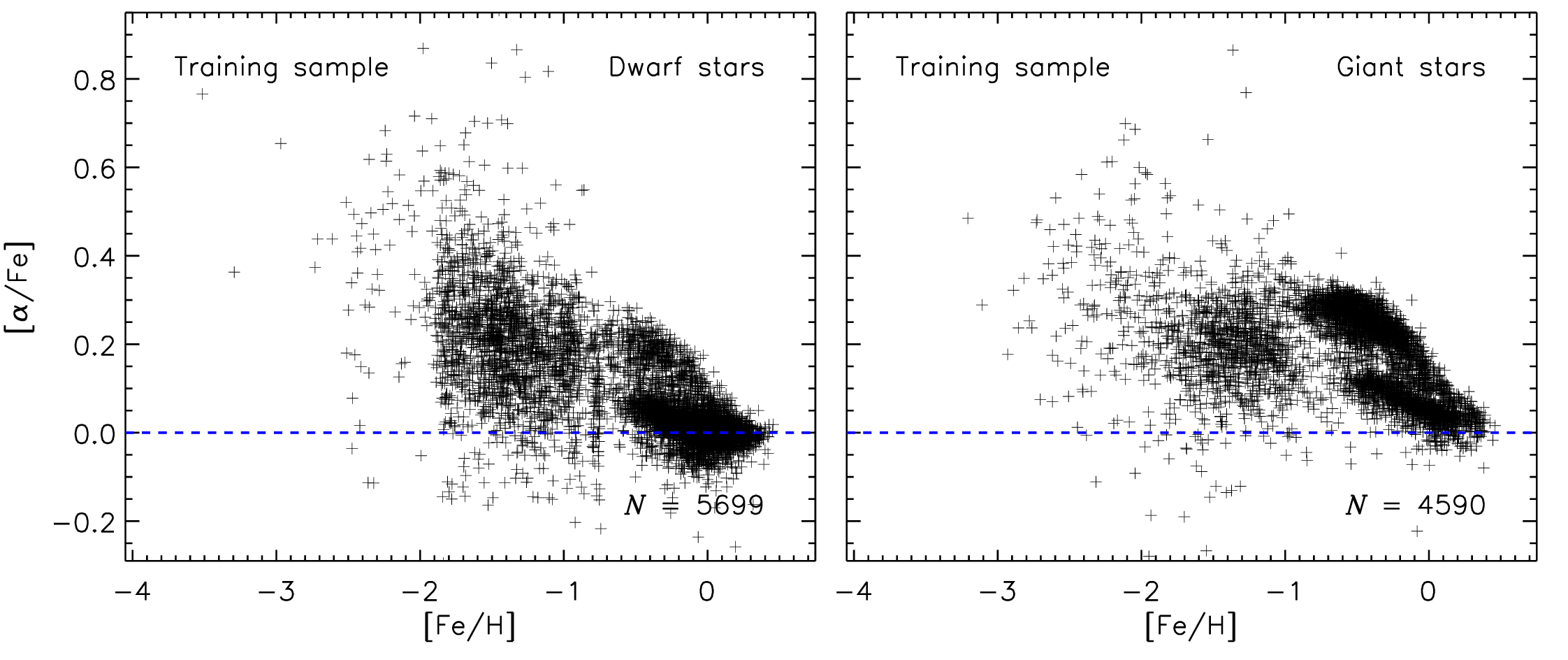}
\includegraphics[scale=0.38,angle=0]{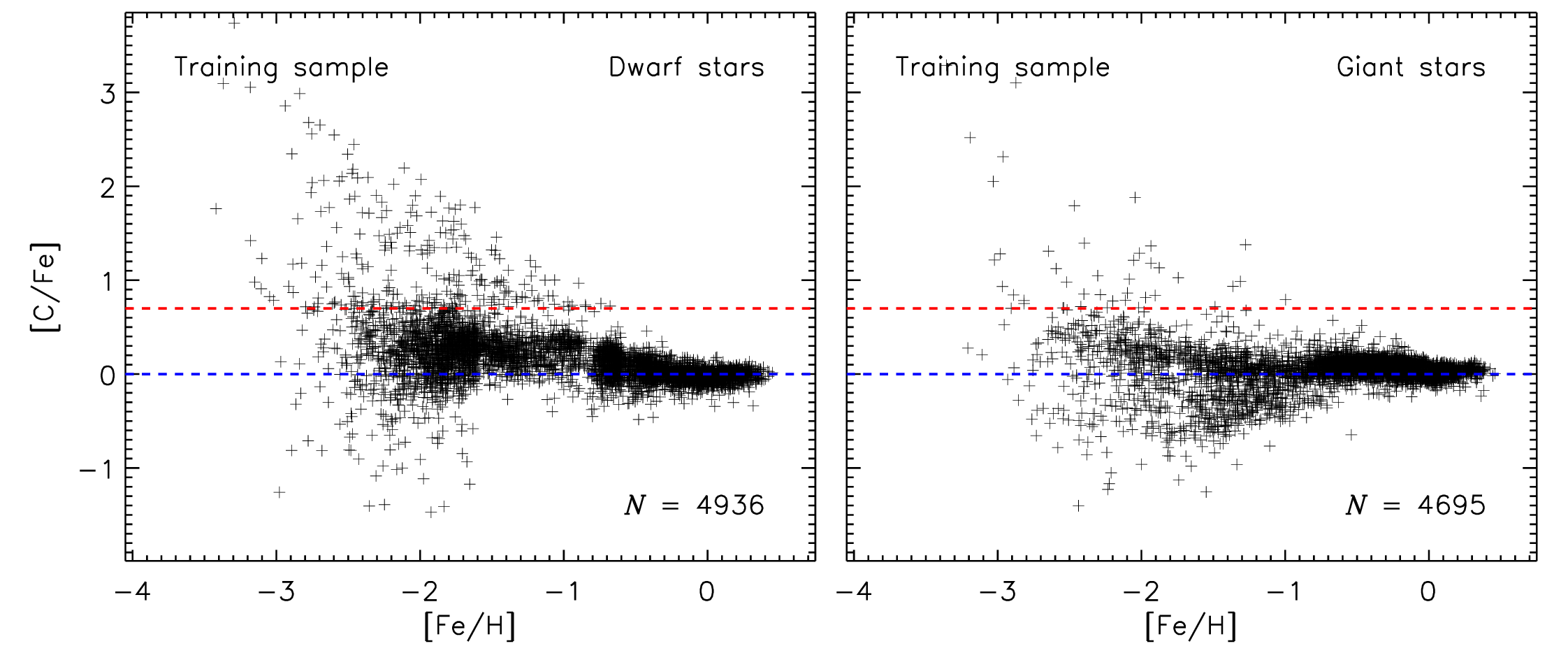}
\caption{Distributions of stars in the training sample: [$\alpha$/Fe] in the [$\alpha$/Fe]–[Fe/H] plane (top panels) and [C/Fe] in the [C/Fe]–[Fe/H] plane (bottom panels). 
The left panels correspond to dwarf stars, while the right panels show giant stars. 
In the bottom panels, the red-dashed lines mark [C/Fe] $= +0.7$, a commonly adopted criterion for defining carbon-enhanced metal-poor (CEMP) stars. 
For the present analysis, we report the measured [C/Fe] values without applying evolutionary corrections. 
The blue-dashed lines in each panel indicate the Solar ratios.}
\label{fig:training_ca}
\end{center}
\end{figure*} 

\begin{figure*}
\begin{center}
\includegraphics[scale=0.3,angle=0]{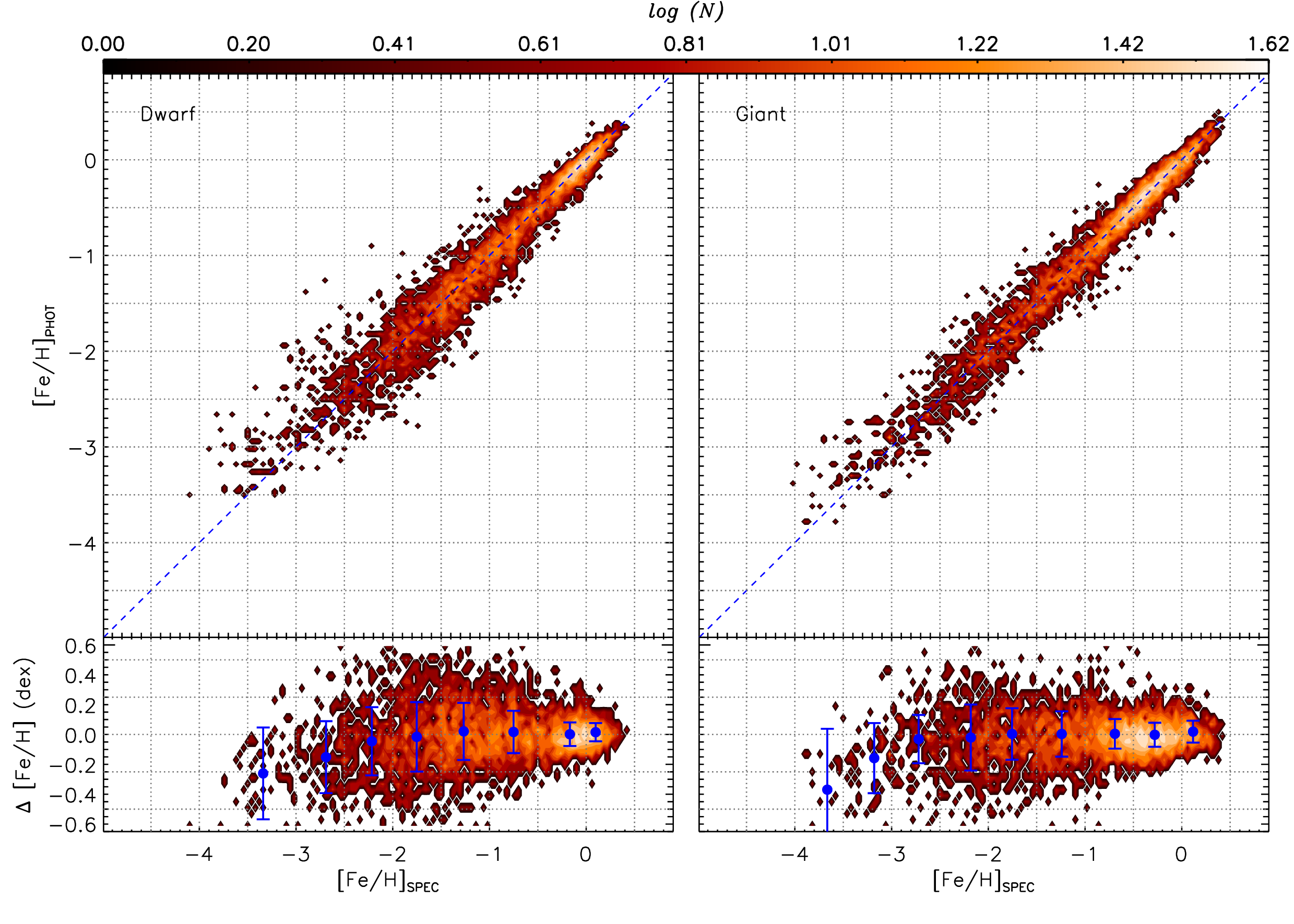}
\caption{Comparisons between photometric and spectroscopic metallicities (top panels) for dwarf stars (left column) and giant stars (right column) in the training sample. 
The photometric metallicities are derived from multiple colors constructed by combining S-PLUS DR4 and $Gaia$ EDR3 magnitudes using the KPCA technique (see text). 
The lower panels show the abundance differences (photometric minus spectroscopic), as a function of the spectroscopic values. 
Blue dots and error bars indicate the median and dispersion of the differences within each bin. 
The blue-dashed lines mark the one-to-one relations, and the color bars at the top indicate the number of stars in each bin.}
\label{fig:training_res1}
\end{center}
\end{figure*} 

\begin{figure*}
\begin{center}
\includegraphics[scale=0.3,angle=0]{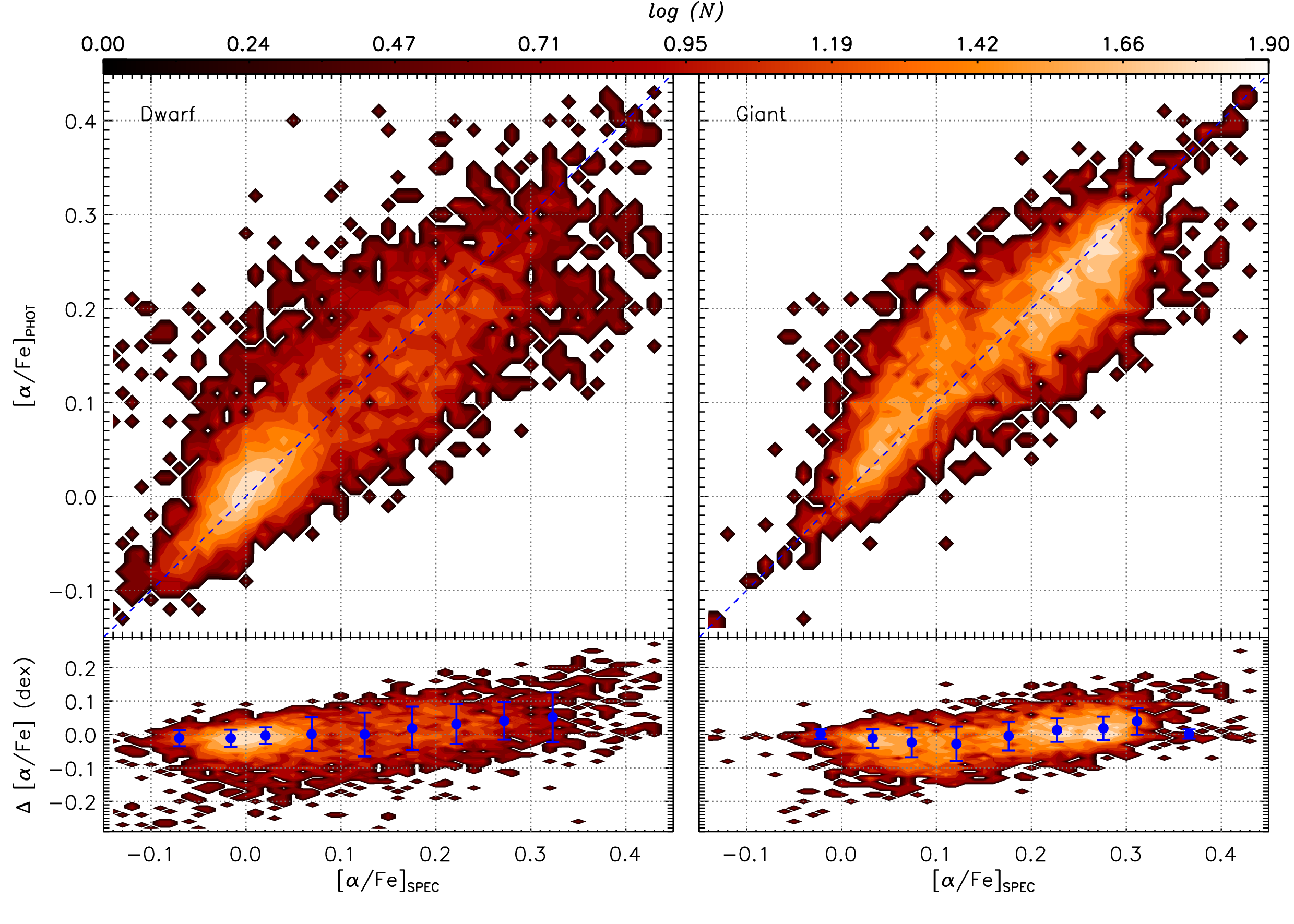}
\includegraphics[scale=0.3,angle=0]{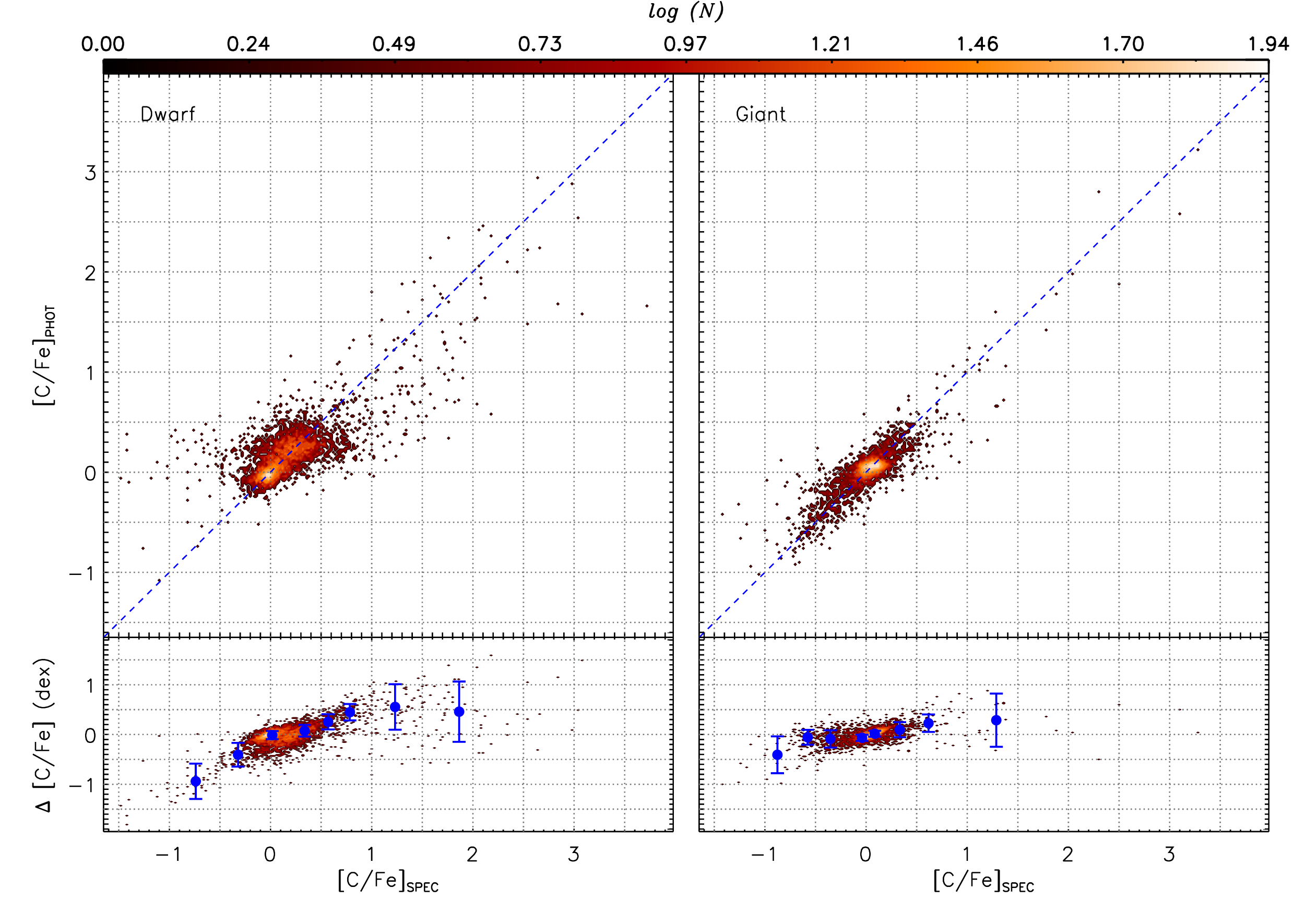}
\caption{Similar to Figure~\ref{fig:training_res1}, but for [$\alpha$/Fe] (top panels) and [C/Fe] (bottom panels).}
\label{fig:training_res2}
\end{center}
\end{figure*}

\begin{figure*}
\begin{center}
\includegraphics[scale=0.225,angle=0]{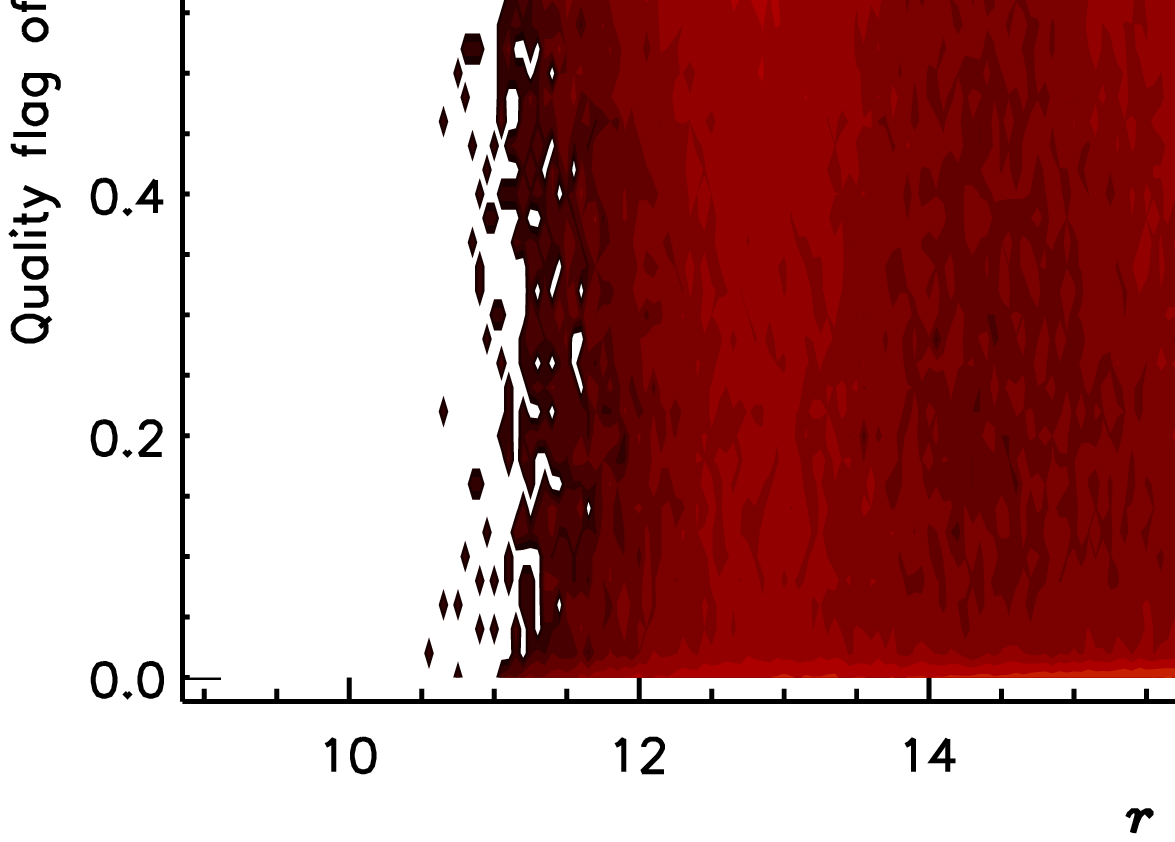}
\includegraphics[scale=0.225,angle=0]{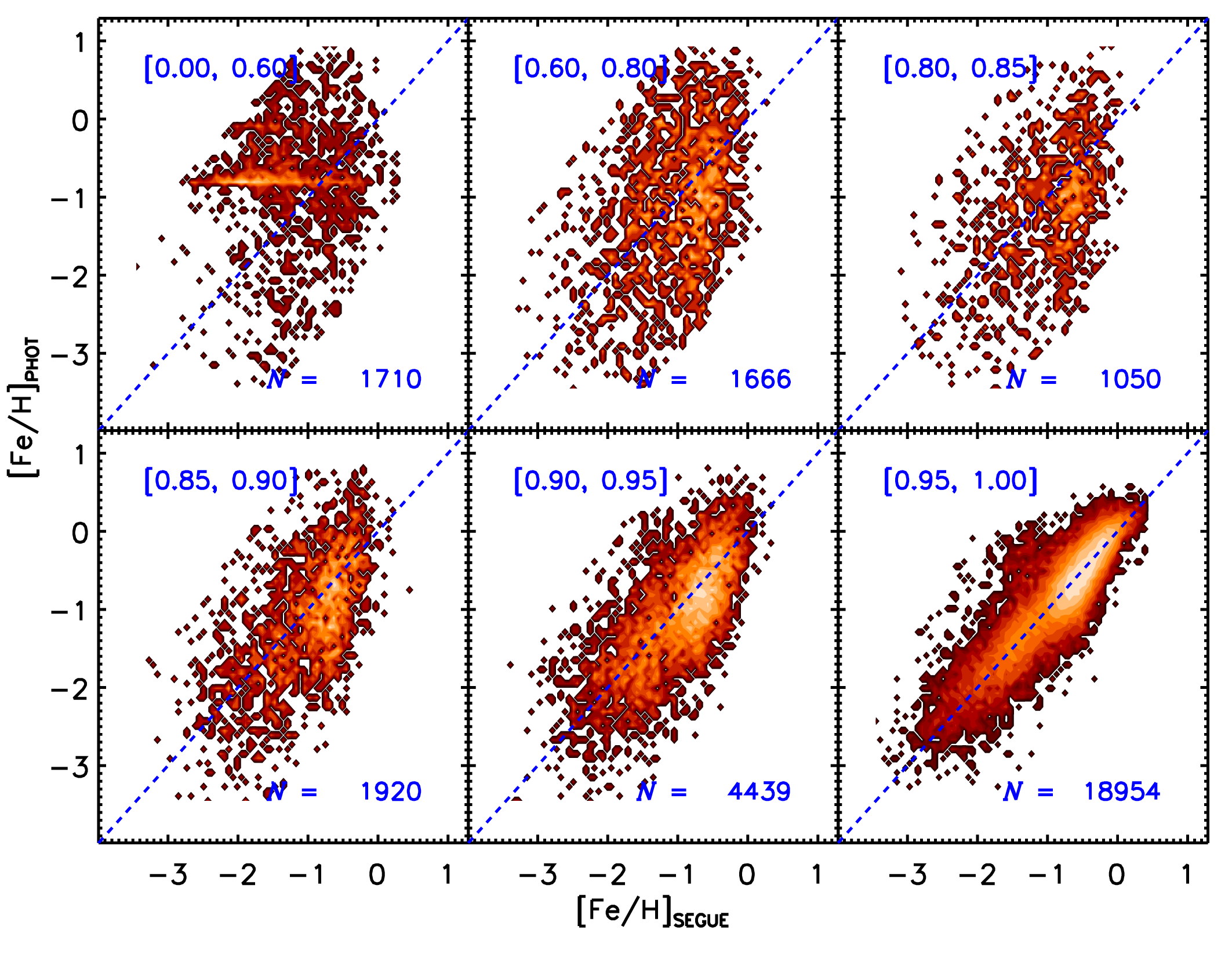}
\caption{Left panel: Density distribution of the [Fe/H] quality flag {\tt flg}$_{\rm [Fe/H]}$, as a function of $r$-band magnitude.  A color bar representing the numbers of stars is provided at the top of the panel.  Right panel: Comparisons of the photometric metallicity and the SDSS/SEGUE spectroscopic metallicity for different ranges of the [Fe/H] quality flag (as marked in the top-left corner of each sub-panel). The blue-dashed lines are the one-to-one lines. The total number of stars used in the comparison is marked in the bottom-right corner of each sub-panel.}
\label{fig:training_res3}
\end{center}
\end{figure*} 

\subsection{Training Sample} \label{sec:training}

The training sample is constructed following the procedure described in Paper III.
In general, the high-resolution spectroscopy (HRS) sample compiled from the PASTEL \citep{2016A&A...591A.118S} and SAGA \citep{2008PASJ...60.1159S}, SDSS/APOGEE \citep{apogee}, LAMOST \citep{2012RAA....12..723Z}, and LAMOST+SEGUE very metal-poor
\citep[VMP with ${\rm [Fe/H]} < -2.0$, derived using the LSSPP developed by][]{2015AJ....150..187L} samples are used for training, while the SDSS/SEGUE and GALAH surveys serve for validation.
All measurements are calibrated to a uniform scale, as detailed in Paper III.

When constructing the training set, we required the stars to meet the following criteria:  
\begin{enumerate}
    \item Photometric uncertainties smaller than 0.04\,mag in the 12 S-PLUS bands and 0.02\,mag in the three \textit{Gaia} bands;  
    \item Moderate reddening, with $E(B-V) \leq 0.10$ and over 90\% of them with $E(B-V) \leq 0.05$;
    \item Metallicities derived from high-quality spectra with a signal-to-noise ratio (SNR) of at least 25 per pixel.  
\end{enumerate}
The selected stars were then separated into dwarfs and giants based on an empirical division in the $(G_{\rm BP} - G_{\rm RP})_0$--$M_{G}$ diagram (see Paper~I). Absolute $G$-band magnitudes were obtained from extinction-corrected \textit{Gaia} $G$-band photometry, and the distances were estimated from \textit{Gaia} parallaxes by \cite{BJ21}.  
All metal-poor stars ([Fe/H] $\lesssim -1.5$) that satisfied these criteria were included. To ensure balanced training across the full metallicity range, a comparable number of more metal-rich stars ([Fe/H] $> -1.5$) were randomly drawn from the large pool of stars meeting the same cuts.  
In total, 4,413 dwarfs and 4,401 giants were selected for the [Fe/H] training sets. Their metallicity distributions are shown in Figure~\ref{fig:training_feh}, extending down to [Fe/H] $\sim -4.0$. Using the same strategy, we obtained 5,699 dwarfs and 4,590 giants for the [$\alpha$/Fe] training sets, and 4,936 dwarfs and 4,695 giants for the [C/Fe] training sets.  
Figure~\ref{fig:training_ca} shows their distributions in the [$\alpha$/Fe]--[Fe/H] and [C/Fe]--[Fe/H] spaces.
 Note that the training set consists exclusively of stars from the main survey regions; stars in the MC regions are not included.

 \begin{table}
\centering
\caption{Summary of the S-PLUS Filter System}
\begin{tabular}{ccccc}
\hline
Filter& $\lambda_{\rm eff}$ &FWHM&$k_{\chi} = \frac{A_{\chi}}{E (B-V)}$&Comments\\
    &(\AA)&(\AA)&&\\
\hline
\hline
$u$&3536&352&4.937&Balmer-break region\\
$J0378$&3770&151&4.664& O II\\
$J0395$&3940&103&4.480& Ca H+K\\
$J0410$&4094&201&4.326& H$\delta$\\
$J0430$&4292&201&4.113&CH $G$-band\\
$g$&4751&1545&3.663&SDSS\\
$J0515$&5133&207&3.334&Mg $b$ triplet\\
$r$&6258&1465&2.515&SDSS\\
$J0660$&6614&147&2.304&H$\alpha$\\
$i$&7690&1506&1.803&SDSS\\
$J0861$&8611&408&1.458& Ca triplet\\
$z$&8813&1182&1.415& SDSS\\
\hline 
\hline
\end{tabular}
\end{table}


\begin{figure*}
\begin{center}
\includegraphics[scale=0.40,angle=0]{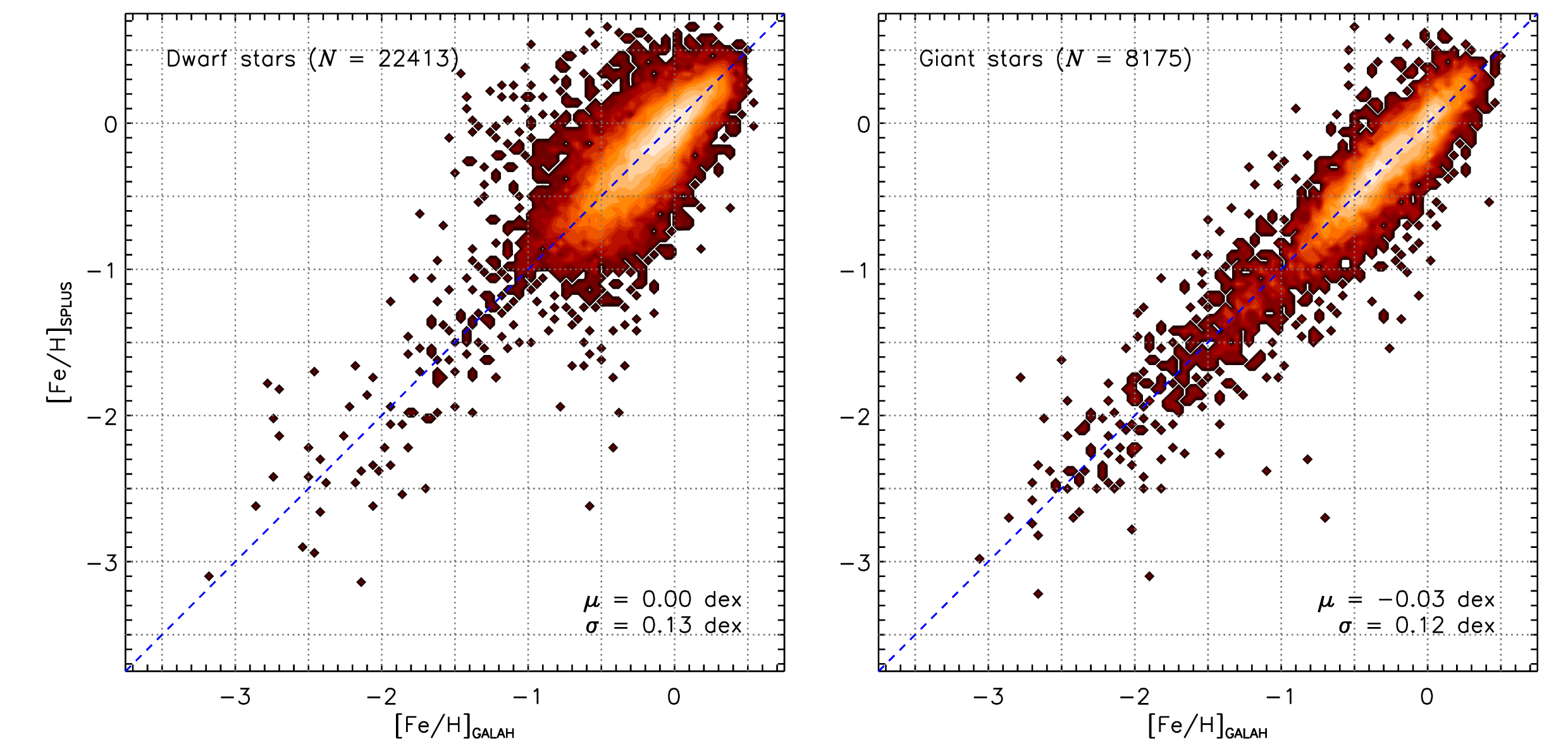}
\includegraphics[scale=0.40,angle=0]{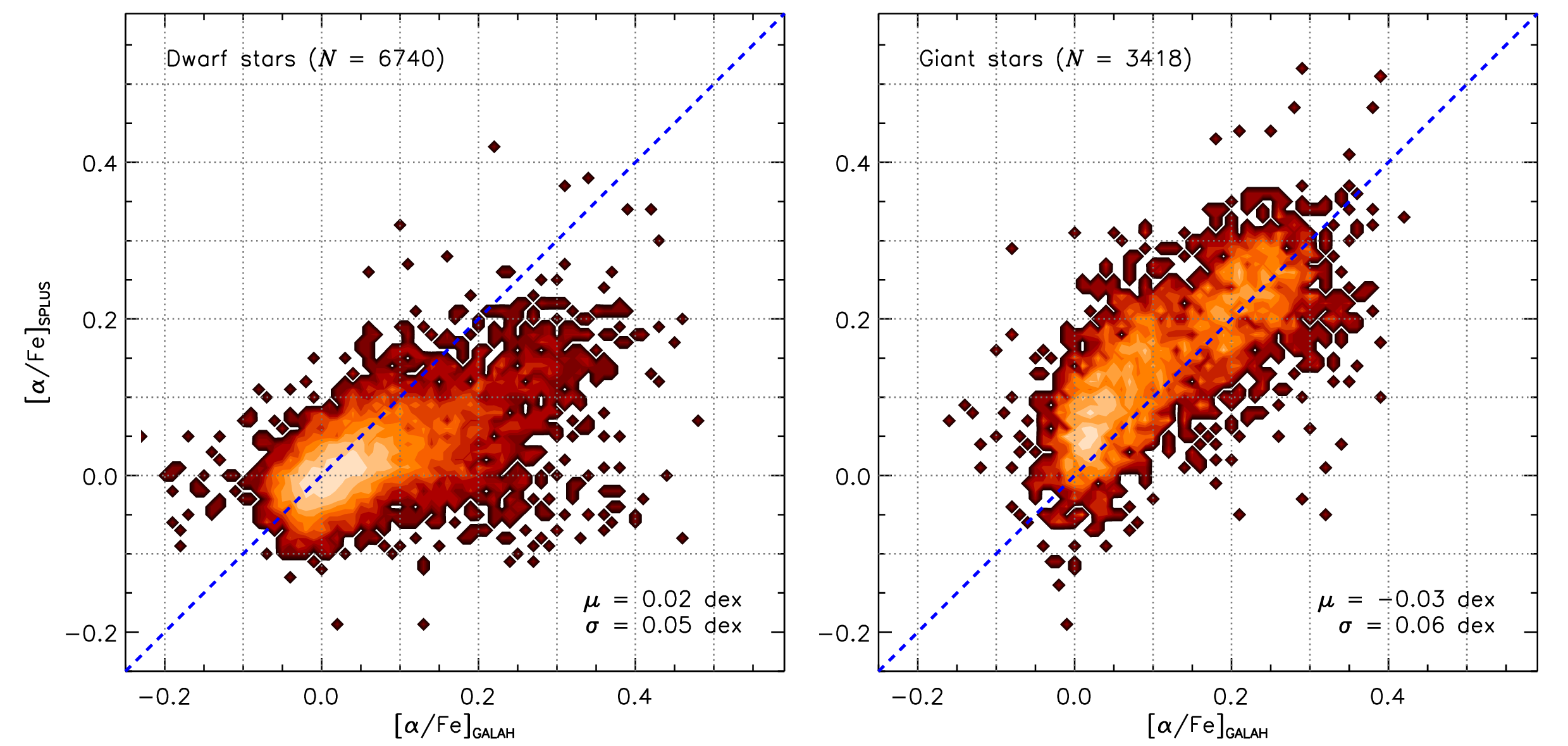}
\caption{Comparisons between our photometric and GALAH DR4 spectroscopic estimates of [Fe/H] (top panels) and [$\alpha$/Fe] (bottom panels). Results for dwarf stars are shown in the left column; for giant stars in the right column. The 
blue-dashed lines indicate the one-to-one relations. The median offset and standard deviation are marked in the the bottom-right corner of each panel.}
\label{fig:comp_galah_feh}
\end{center}
\end{figure*}

\begin{figure*}
\begin{center}
\includegraphics[scale=0.305,angle=0]{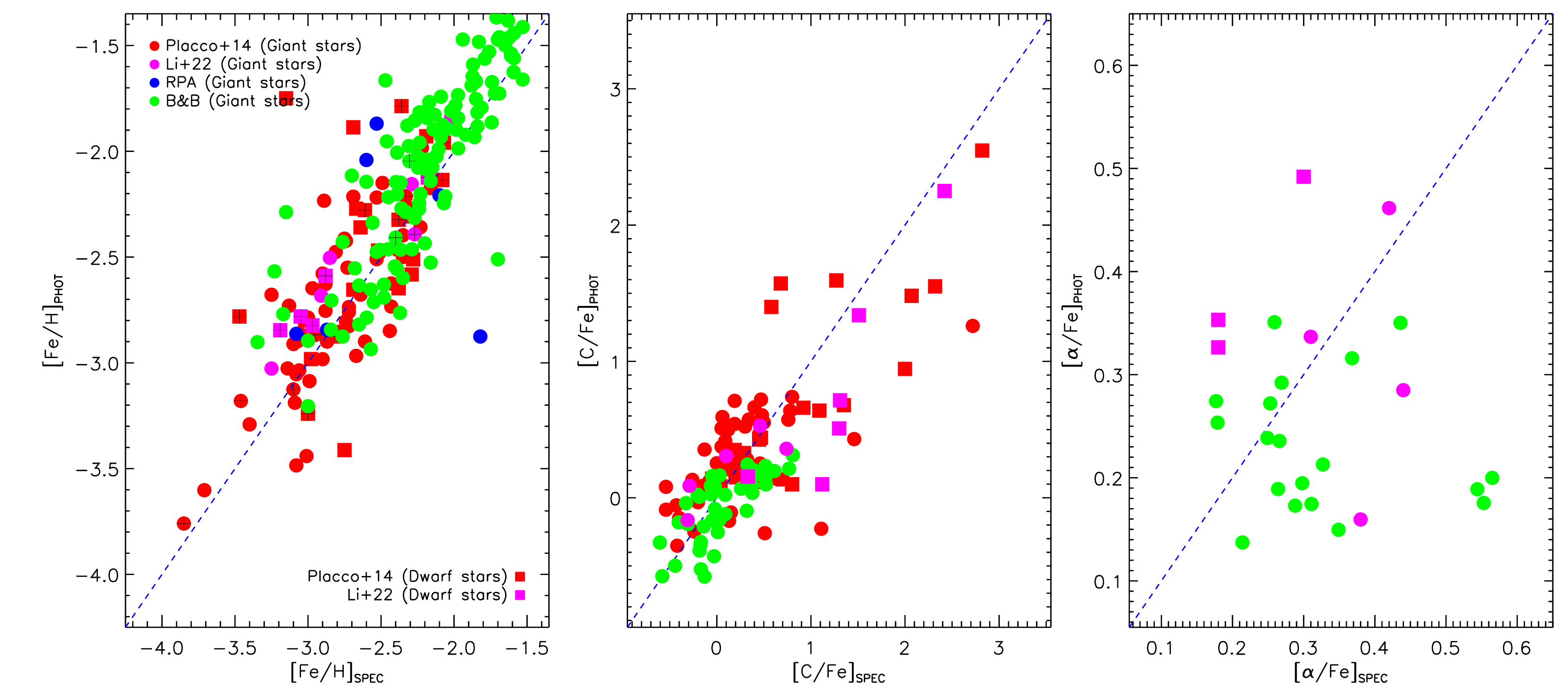}
\caption{Comparisons of the photometric estimates of [Fe/H] (left), [C/Fe] (middle), and [$\alpha$/Fe] (right) with spectroscopic measurements from several samples: the Best \& Brightest Survey \citep[green dots;][]{BNB}, high-resolution spectroscopic (HRS) samples, CEMP stars from \citet[][red dots]{Placco14}, the $R$-Process Alliance \citep[blue dots;][]{2018ApJ...858...92H, 2018ApJ...868..110S, 2020ApJ...898..150E, 2020ApJS..249...30H}, and Subaru follow-up observations of LAMOST VMP candidates \citep[magenta dots;][]{2022ApJ...931..146A, 2022ApJ...931..147L}. In the [Fe/H] panel, plus symbols indicate CEMP stars with [C/Fe]\,$> +0.7$. The blue-dashed lines represent the one-to-one relations.}
\label{fig:comp_lit}
\end{center}
\end{figure*} 

\begin{figure*}
\begin{center}
\includegraphics[scale=0.28,angle=0]{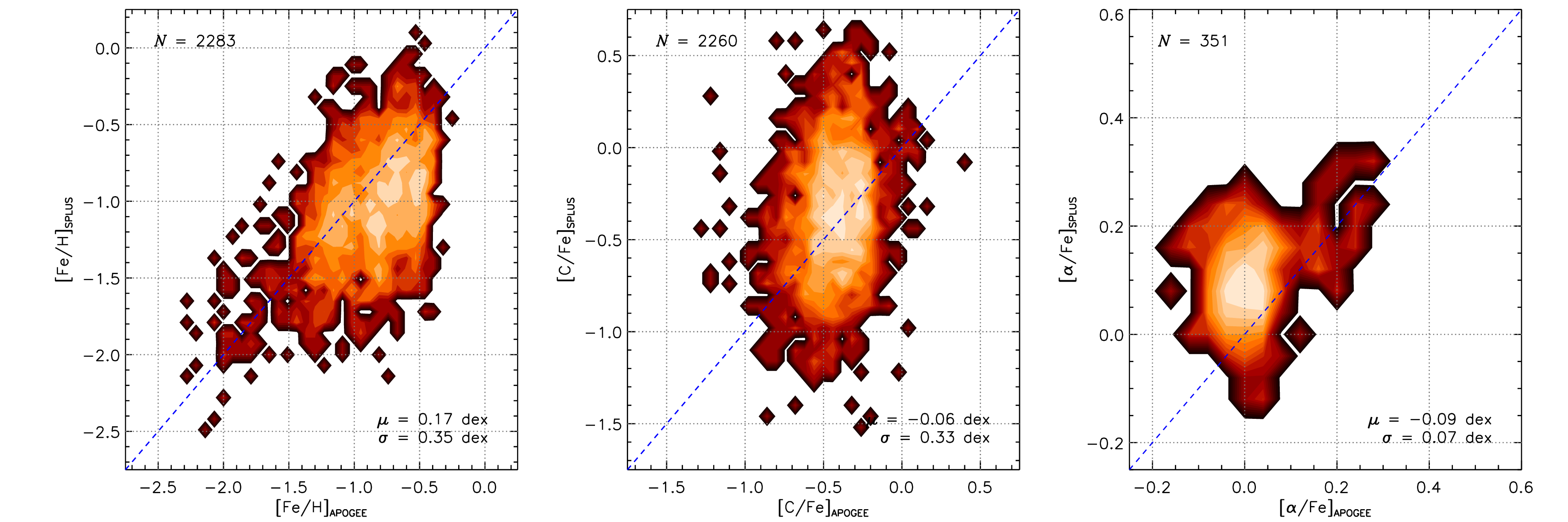}
\caption{Comparisons of the photometric estimates with APOGEE DR17 spectroscopic values for MC member stars: [Fe/H] (left panel), [C/Fe] (middle panel), and [$\alpha$/Fe] (right panel). The blue-dashed lines represent the one-to-one relations. The median offset and standard deviation are indicated in the bottom-right corner of each panel.}
\label{fig:comp_mc}
\end{center}
\end{figure*} 

\begin{figure*}
\begin{center}
\includegraphics[scale=0.355,angle=0]{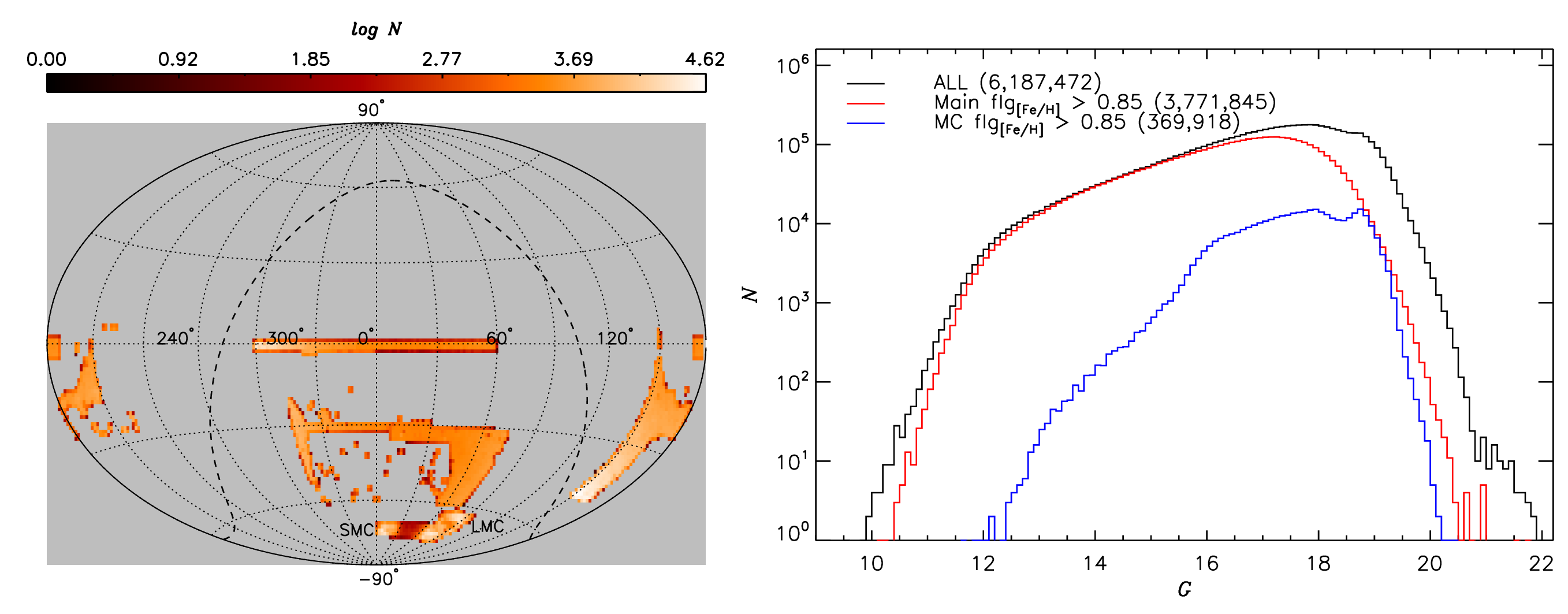}
\caption{Left panel: Density distribution of stars in the final S-PLUS parameter sample across the sky in equatorial coordinates. The black-dashed line indicates the Galactic plane.
Positions of the LMC and SMC are marked.
Right panel: Magnitude distribution of stars in the final S-PLUS parameter sample in the \textit{Gaia} $G$ band. The black, red, and blue histograms correspond to all sample stars with {\tt flg$_{\rm [Fe/H]} \ge 0.60$}, main-survey stars, and MC stars with {\tt flg$_{\rm [Fe/H]} \ge 0.85$}, respectively.}
\label{fig:distri}
\end{center}
\end{figure*} 

\begin{figure*}
\begin{center}
\includegraphics[scale=0.3,angle=0]{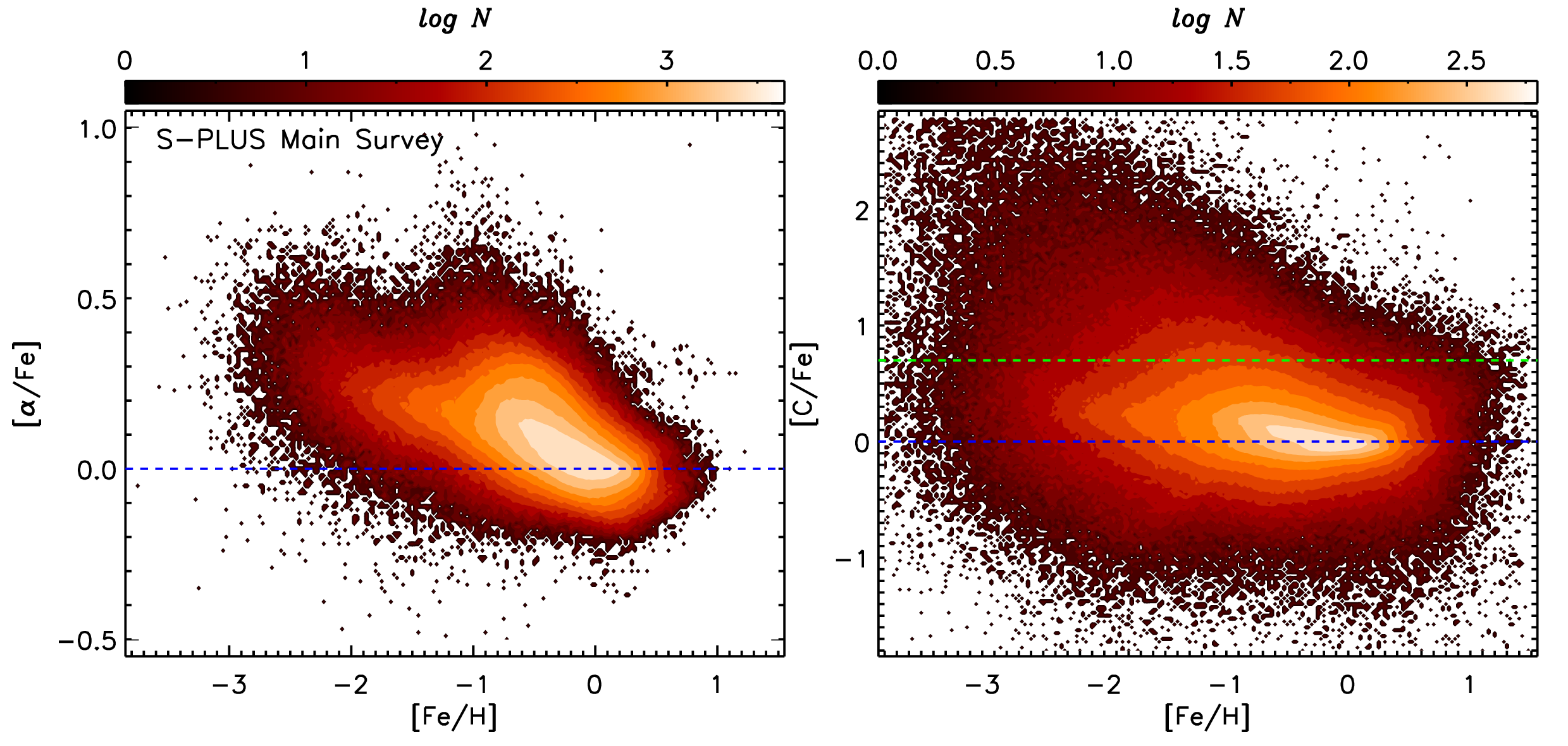}
\includegraphics[scale=0.3,angle=0]{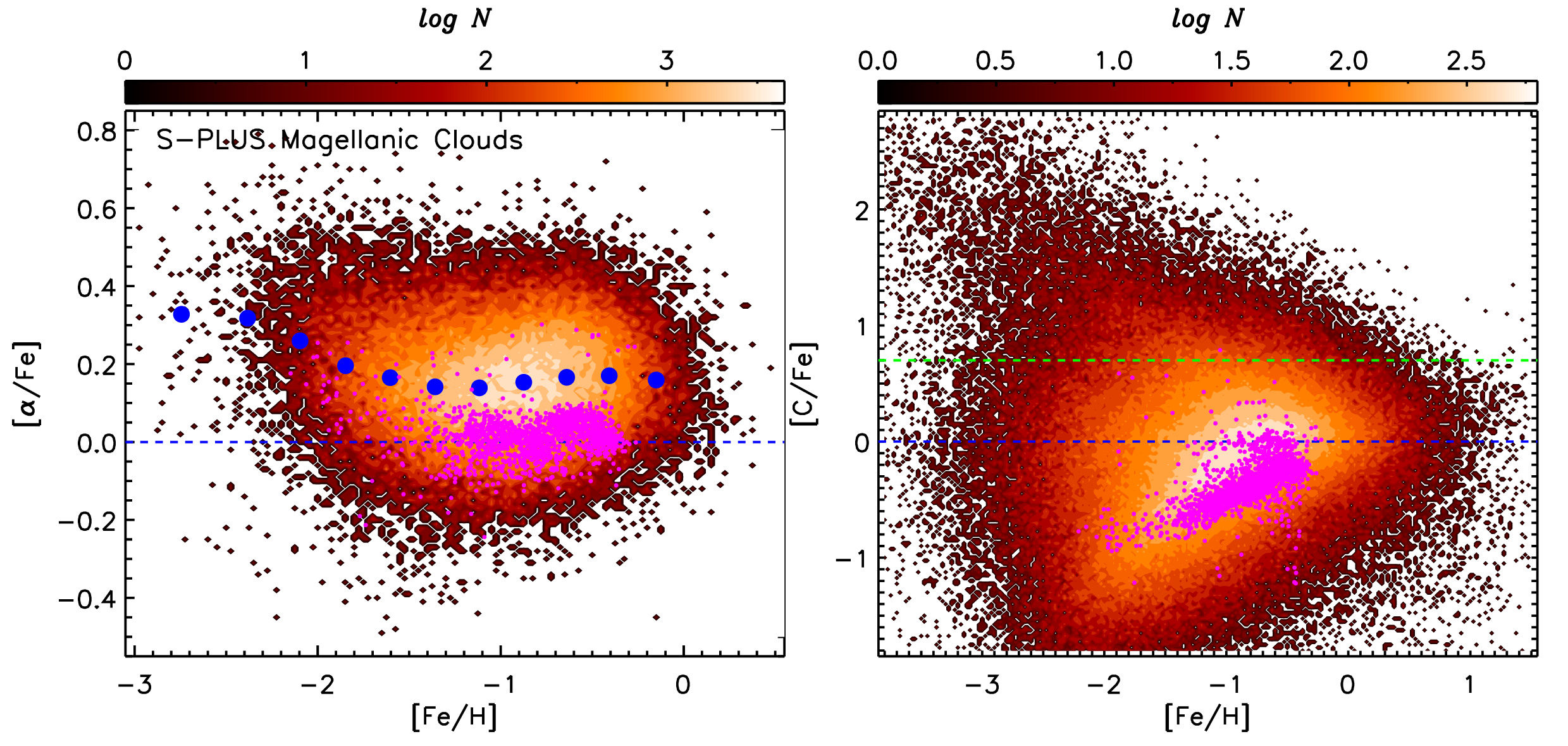}
\caption{Upper panels: Density distributions of [$\alpha$/Fe] versus [Fe/H] (left) and [C/Fe] versus [Fe/H] (right) for main-survey stars with 
{\tt flg$_{\rm [Fe/H]} \ge 0.85$}, 
{\tt flg$_{\rm [\alpha/Fe]} \ge 0.85$}, and 
{\tt flg$_{\rm [C/Fe]} \ge 0.85$}. 
A color bar is shown at the top of each panel.  
Lower panels: Same as the upper panels, but for MC member stars. 
The blue-dashed lines indicate the Solar ratios in each panel, 
while the green-dashed line marks [C/Fe] = +0.70. 
Magenta dots represent stars in common with APOGEE DR17, with their abundance measurements taken from APOGEE DR17.
The blue dots mark the mean [$\alpha$/Fe] values in individual [Fe/H] bins.
These blue dots exhibit a trend in [$\alpha$/Fe] versus [Fe/H] similar to that seen for stars in common with APOGEE DR17, where elemental abundances were measured spectroscopically, although a moderate offset is present, as already shown in Figure~\ref{fig:comp_mc}.}
\label{fig:distri_param}
\end{center}
\end{figure*}

\begin{figure*}
\begin{center}
\includegraphics[scale=0.305,angle=0]{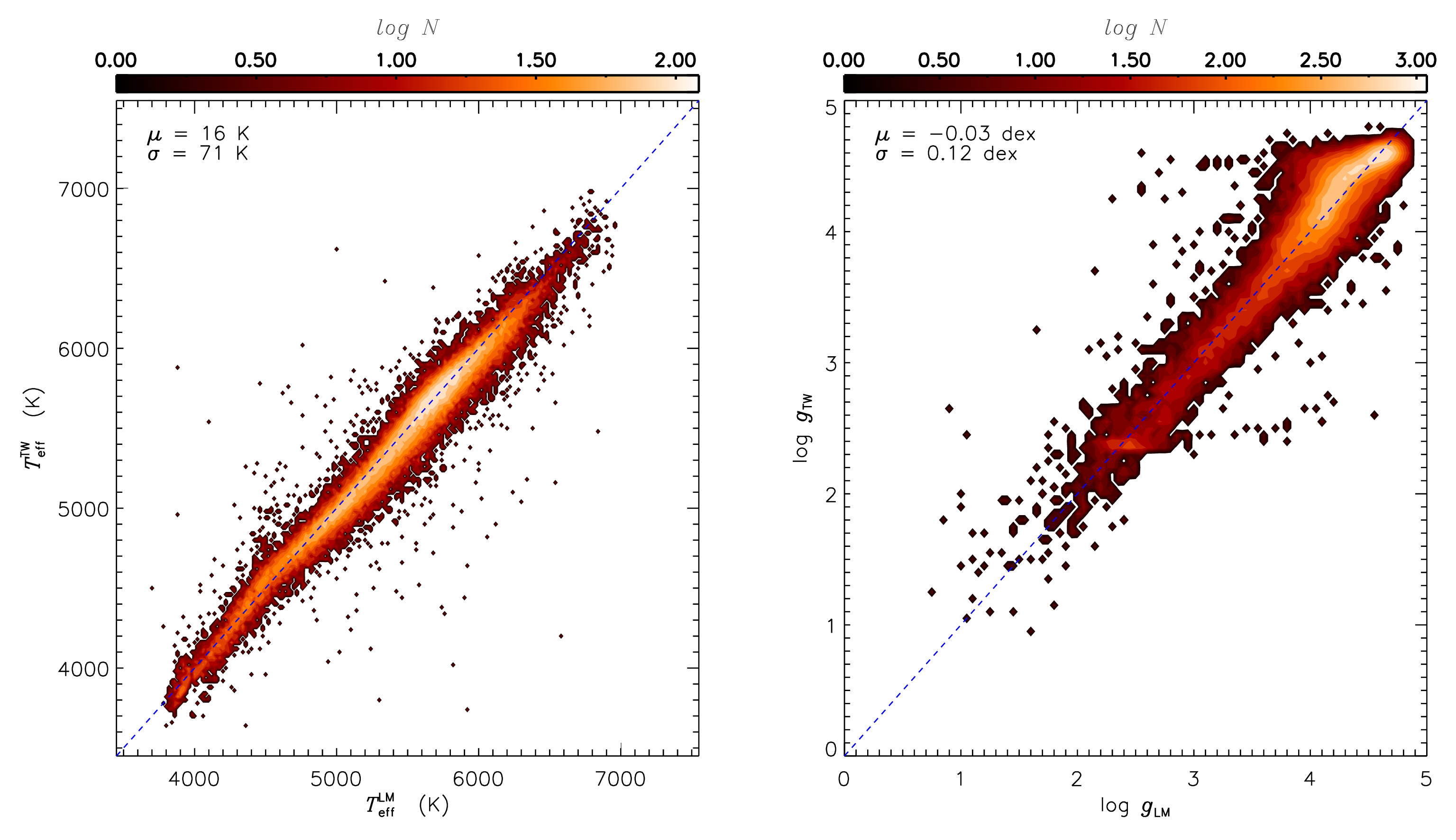}
\caption{Comparisons between our estimates of $T_{\rm eff}$ (left panel) and log~$g$ (right panel) and those from DR11 of the LAMOST low-resolution survey. The blue-dashed lines indicate the one-to-one relations. The overall median offset and standard deviation are shown in the top-left corner of each panel. Color bars at the top indicate the number density of stars.}
\label{fig:comp_tg}
\end{center}
\end{figure*} 
\section{Estimates of [Fe/H], [C/Fe], and [$\alpha$/Fe] from the S-PLUS and $Gaia$ Colors}

\subsection{Training Results} \label{sec:training_res}
Following the same procedures employed in Paper~III, we trained relations between metallicity (or elemental abundances) and 13 stellar colors, $G_{\rm BP} - G_{\rm RP}$, $G_{\rm BP} - u$, $G_{\rm BP} - g$, $G_{\rm RP} - r$, $G_{\rm RP} - i$, $G_{\rm RP} - z$, $G_{\rm BP} - J0378$, $G_{\rm BP} - J0395$, $G_{\rm BP} - J0410$, $G_{\rm BP} - J0430$, $G_{\rm BP} - J0515$, $G_{\rm RP} - J0660$, and $G_{\rm RP} - J0861$, using the KPCA method \citep{S99}. All colors were
de-reddened with the SFD extinction map and the extinction coefficients described in Section~\ref{sec:GEDR3}.

The training result for [Fe/H] is presented in Figure~\ref{fig:training_res1}, while those for [$\alpha$/Fe] and [C/Fe] are shown in Figure~\ref{fig:training_res2}. Overall, the photometric estimates of metallicity are consistent with the spectroscopic measurements, even at the extremely metal-poor end down to [Fe/H]~$\sim -4.0$. No significant offsets are seen down to [Fe/H]$=-3.0$, though small deviations ($-0.10$ to $-0.30$\,dex; spectroscopic minus photometric) appear at metallicities lower than $-3.0$. The dispersion is very small ($<0.10$\,dex) in the relatively metal-rich regime ([Fe/H] $>-1.0$) and increases to 0.10--0.30\,dex in the metal-poor regime ([Fe/H]~$<-1.0$). We also find that the [Fe/H] precision for giant stars is generally better than that for dwarfs, particularly at low metallicities, as expected because of the weaker metallic absorption lines in warmer dwarfs.

For [$\alpha$/Fe], the photometric estimates agree well with the spectroscopic measurements, achieving a precision better than 0.05\,dex for both dwarfs and giants in the low-$\alpha$ regime. In the high-$\alpha$ regime, slight systematic offsets appear, and the uncertainties increase to 0.05–0.10\,dex.
For [C/Fe], the photometric estimates for dwarf stars exhibit moderate offsets in both the carbon-rich ([C/Fe] $> +1.0$; around $+0.5$\,dex, spectroscopic minus photometric) and carbon-poor ([C/Fe] $< -0.5$; around $-0.8$\,dex) regimes. The precision depends on [C/Fe]: it is 0.05--0.15\,dex in the intermediate range ($-0.5 <$ [C/Fe] $< +0.5$), but degrades to 0.2--0.4\,dex toward the carbon-rich and carbon-poor ends. For giant stars, the photometric [C/Fe] estimates generally agree well with the spectroscopic values, with moderate offsets of 0.20--0.40\,dex only at the extreme ends. The precision is 0.05--0.10\,dex in the intermediate region and 0.15--0.30\,dex at the carbon-rich and carbon-poor ends.

\subsection{Application to the S-PLUS Parent Sample}

We apply the trained relationships to 8 million stars from the main survey and an additional 2 million stars in the MC regions. 
As in Paper~III, we introduce a quality parameter, {\tt flg$_x$} (where $x=$ [Fe/H], [C/Fe] or [$\alpha$/Fe]), to assess the reliability of the abundance estimates. 
This parameter is defined as the maximum kernel value between the target stellar colors and those in the training set. 
Its value ranges from 0 to 1, with unity indicating an exact match between the colors of a target and those of the training set. As an illustration, Figure~\ref{fig:training_res3} shows the distribution of {\tt flg$_{\rm [Fe/H]}$} as a function of the $r$-band magnitude. As expected, {\tt flg$_{\rm [Fe/H]}$} remains close to unity for stars brighter than $r=18.5$, but decreases rapidly toward zero at the faint end owing to larger photometric errors. 
We note that a small fraction of bright stars also exhibit low values of {\tt flg$_{\rm [Fe/H]}$}, likely caused by saturation, stellar variability, binarity, emission-line features, or other peculiarities.

We compare S-PLUS photometric metallicities with SDSS/SEGUE spectroscopy\footnote{We note that the SEGUE VMP stars included in the training sample have been excluded from this analysis.} in bins of {\tt flg$_{\rm [Fe/H]}$} (see the right panel of Figure~\ref{fig:training_res3}). The dispersion between photometric and spectroscopic values increases as {\tt flg$_{\rm [Fe/H]}$} decreases. For {\tt flg$_{\rm [Fe/H]}$} $< 0.60$, the dispersion exceeds 1.0\,dex and artificial features become apparent; thus, photometric metallicities are considered unreliable below this threshold.
After excluding stars used in the training sets, the remaining APOGEE–S-PLUS stars in common are used to examine abundance differences (spectroscopic minus photometric) as a function of {\tt flg$_x$}. 
The resulting {\tt flg$_x$}-dependent dispersion, $\sigma_x$, is then adopted as the random error for each star. Additionally, the uncertainty of the method for different elemental abundance bins can be estimated from the training results, as shown in Figure~\ref{fig:training_res1}. Accordingly, the final uncertainty of an abundance measurement for a given star is calculated as 
\[
\sigma_{\rm final} = \sqrt{\sigma_x^2 + \sigma_m^2},
\] 
where $\sigma_x$ represents the random error and $\sigma_m$ the method error.
The calculation of $\sigma_x$ is described above, while the values of $\sigma_m$ depend on the abundances themselves, as shown in Figures~\ref{fig:training_res1} and~\ref{fig:training_res2}.
For a detailed description of the uncertainty calculation procedure, we refer the interested reader to Paper~III.

In total, 5.42 million stars from the main survey and 0.76 million stars in the MC regions have abundance determinations with at least one quality parameter ({\tt flg$_x$}) exceeding 0.6. These stars are hereafter referred to as the S-PLUS parameter sample.
We emphasize that for each parameter type $x$, where $x$ represents [Fe/H], [$\alpha$/Fe], or [C/Fe], a cut of {\tt flg$_x > 0.6$} must be applied.

\subsection{Validation}
\subsubsection{Comparison with GALAH DR4}

As noted in Section~\ref{sec:training}, the GALAH survey stars are not included in the training sets and can therefore be used to independently test our photometric estimates. 
We cross-match GALAH DR4 \citep{2025PASA...42...51B} with the S-PLUS parameter sample, requiring GALAH {\tt SNR\_C2\_IRAF}\,$\geq 30$ for [Fe/H] or $\geq 50$ for [$\alpha$/Fe] per pixel, and S-PLUS {\tt flg$_x$}\,$\geq 0.9$. 
The comparison results are presented in Figure~\ref{fig:comp_galah_feh}. 
For [Fe/H], our photometric estimates exhibit excellent agreement with the high-resolution spectroscopic values from GALAH, with negligible offsets and a scatter of about 0.10\,dex for both dwarf and giant stars. 
Similarly, the photometric estimates of [$\alpha$/Fe] are quite precise, with scatters of only 0.05\,dex for dwarfs and 0.06\,dex for giants. However, we note that the photometric results for high-[$\alpha$/Fe] stars are less reliable. These results are consistent with the training performance shown in Figure~\ref{fig:training_res1}.

\subsubsection{Comparison with Metal-poor Samples from the Literature}

First, we compare our photometric metallicities with those of low-metallicity candidates from the Best \& Brightest (B\&B) Survey, which have been followed up with low/medium-resolution spectroscopy ($R \sim 2000$) \citep{BNB}. 
More recently, \citet{2019ApJ...870..122P} and \citet{Limberg21} presented metallicity estimates, as well as other elemental-abundance ratios ([C/Fe] and [Mg/Fe]), for nearly 1900 stars using the low/medium-resolution spectra obtained from the B\&B follow-up \footnote{We note that the spectroscopic abundance ratios for [C/Fe], [Mg/Fe], or [$\alpha$/Fe] reported by \citet{2019ApJ...870..122P} and \citet{Limberg21} have been superseded by the values listed in \citet{2022ApJ...926...26S}.}.

In total, we identify 131 metal-poor ([Fe/H]\,$\leq -1.5$) giants in common, with 
{\tt flg$_{\rm [Fe/H]}$}\,$\geq 0.9$, including stars with metallicities as low as 
[Fe/H] $= -3.0$. The metallicity comparison exhibits a moderate offset of $-0.13$\,dex 
(B\&B minus photometric estimates), as illustrated in Figure~\ref{fig:comp_lit}, with 
a scatter of 0.24\,dex. This level of scatter is consistent with expectations from 
our internal tests (Figure~\ref{fig:training_res1}; see Section~\ref{sec:training_res}).  
For [C/Fe], there are 41 stars in common with reliable B\&B measurements and 
{\tt flg$_{\rm [C/Fe]}$}\,$\geq 0.9$ in our catalog. In general, the photometric 
[C/Fe] estimates\footnote{The [C/Fe] estimates refer to results without 
evolution-dependent corrections, since no such corrections were applied to the 
training sets (see Section~\ref{sec:training}).} are in good agreement with those 
from the B\&B follow-up, with a moderate offset of $+0.18$\,dex and a scatter of 
0.25\,dex.  
For [$\alpha$/Fe], the scatter in the differences is larger. This may be due to 
(1) the relatively low precision of both sets of measurements and 
(2) the small number of stars in common (only 21 with reliable measurements in both catalogs).

Secondly, we compare our photometric abundances with those of the HRS sample, 
which are not included in our training sets. 
The HRS sample is compiled from several sources: the CEMP sample with over 600 stars 
\citep{Placco14}; the $R$-Process Alliance project, which includes more than 600 VMP stars 
\citep{2018ApJ...858...92H, 2018ApJ...868..110S, 2020ApJ...898..150E, 2020ApJS..249...30H}; 
and $\sim 400$ LAMOST-selected VMP candidates with chemical abundances derived from 
Subaru high-resolution spectroscopic follow-up 
\citep{2022ApJ...931..146A, 2022ApJ...931..147L}.  
In total, 94 stars (28 dwarfs and 66 giants) are in common with 
{\tt flg$_{\rm [Fe/H]}$}\,$> 0.9$, covering a metallicity range of 
[Fe/H] $= [-4.0, -2.0]$. For [Fe/H], the median offset is moderate 
($-0.14$\,dex; spectroscopy minus photometry) for dwarfs and small 
($-0.09$\,dex) for giants, with scatters of 0.26\,dex and 0.24\,dex, respectively. 
These results are consistent with the performance of the training tests 
(see Figure~\ref{fig:training_res1}).  
As noted in Paper~III, the carbon-sensitive band centered on the CH $G$-band 
($J0430$ filter) enables a direct measurement of carbon abundance, thus helping 
to break the degeneracy between metallicity and carbon enrichment. 
Among the 86 stars in common (27 dwarfs and 59 giants) with 
{\tt flg$_{\rm [C/Fe]}$}\,$\geq 0.9$, nearly all CEMP stars 
([C/Fe]\,$> +0.7$) in the HRS sample are successfully recovered by our photometric estimates. 
The median offset (HRS minus photometric) of [C/Fe] is moderate, $+0.23$\,dex for dwarfs and $-0.20$\,dex for giants, consistent with the trends observed in our training tests (see the bottom panels of Figure~\ref{fig:training_res2}). 
The overall scatter in the [C/Fe] differences is 0.41\,dex for dwarfs and 0.22\,dex for giants.
The scatter between the photometric and HRS [$\alpha$/Fe] estimates remains large, owing in part to the small number of stars in common.

\begin{table*}
\centering
\caption{Description of the Final Sample}
\begin{tabular}{lll}
\hline
\hline
Field&Description&Unit\\
\hline
Sourceid$^*$&Gaia EDR3 source ID&--\\
ra$^*$&Right Ascension from J-PLUS DR3 (J2000)&degrees\\
dec$^*$&Declination from J-PLUS DR3 (J2000)&degrees\\
G/BP/RP$^*$&Magnitudes for the {\it Gaia} three bands from EDR3; note G represents a calibration-corrected G magnitude&--\\
err\_G/BP/RP$^*$&Uncertainties of magnitudes for the three {\it Gaia} bands from EDR3&mag\\
ebv\_sfd&Value of $E (B - V)$ from the extinction map of SFD98, corrected for a 14\% systematic&--\\
\text{[Fe/H]}$^*$&Photometric metallicity&--\\
\text{err\_[Fe/H]}$^*$&Uncertainty of photometric metallicity&dex\\
\text{flg\_[Fe/H]}$^*$&Quality flag of [Fe/H]&--\\
\text{[C/Fe]}$^*$&Photometric carbon-to-iron abundance ratio&--\\
\text{err\_[C/Fe]}$^*$&Uncertainty of photometric carbon-to-iron abundance ratio&dex\\
\text{flg\_[C/Fe]}$^*$&Quality flag of [C/Fe]\\
\text{[$\alpha$/Fe]}$^*$&Photometric alpha-to-iron abundance ratio&--\\
\text{err\_[$\alpha$/Fe]}$^*$&Uncertainty of photometric alpha-to-iron abundance ratio &dex\\
\text{flg\_[$\alpha$/Fe]}$^*$&Quality flag of [$\alpha$/Fe]&-- \\
$T_{\rm eff}$$^*$&Effective temperature&K\\
err\_$T_{\rm eff}$$^*$&Uncertainty of effective temperature&K\\
log\,$g$&Surface gravity&--\\
err\_${\rm log} g$&Uncertainty of surface gravity&dex\\
dist&Distance&pc\\
err\_dist&Uncertainty of distance&pc\\
flg\_dist&Flag to indicate the method used to derive distance, which takes the values ``GaiaEDR3", ``CMD", and ``NO"&--\\
\text{age}&Stellar age&Gyr\\
\text{err\_age}&Uncertainty of stellar age&Gyr\\
rv&Radial velocity&km s$^{-1}$\\
err\_rv&Uncertainty of radial velocity&km s$^{-1}$\\
flg\_rv&Flag to indicate the source of radial velocity, which takes the values ``GALAH-DR3", ``APOGEE-DR17'',&--\\
& ``Gaia-DR3",``RAVE-DR5", ``LAMOST-DR11", ``SDSS-DR12"&--\\
parallax&Parallax from {\it Gaia} EDR3&mas\\
err\_parallax&Uncertainty of parallax from {\it Gaia} EDR3&mas\\
pmra$^*$&Proper motion in Right Ascension direction from  {\it Gaia} EDR3&mas yr$^{-1}$\\
err\_pmra$^*$&Uncertainty of proper motion in Right Ascension direction from  {\it Gaia} EDR3&mas yr$^{-1}$\\
pmdec$^*$&Proper motion in Declination direction  from  {\it Gaia} EDR3&mas yr$^{-1}$\\
err\_pmdec$^*$&Uncertainty of proper motion in Declination direction from  {\it Gaia} EDR3&mas yr$^{-1}$\\
ruwe$^*$&Renormalised unit weight error from  {\it Gaia} EDR3&--\\
type&Flag to indicate classifications of stars, which takes the values ``dwarf" and ``giant" &--\\
subtype&Flag to indicate further sub-classifications of dwarf stars, which takes the values  ``TO", ``MS" and ``Binary"&--\\
\hline
\end{tabular}
\vspace{2mm}
{\textbf{Note.} For stars in the Magellanic Clouds, only parameters marked with $^{*}$ are available.}
\end{table*}

\subsubsection{Performance of Stellar-Parameter Estimates for MC Member Stars}
In addition to the main survey, we derive stellar parameters for more than 0.7 million red giant stars in the MC regions, identified as members based on their $Gaia$ DR3 proper motions (see Appendix). To evaluate the performance of the photometric parameters, we cross-matched these stars with APOGEE DR17 \citep{2022ApJS..259...35A}. For stars in common, we further require an APOGEE SNR $>50$ and at least one reliable photometric quality flag (\texttt{flg}$_{\rm [Fe/H]}\geq0.95$, \texttt{flg}$_{\rm [C/Fe]}\geq0.95$, or \texttt{flg}$_{\rm [\alpha/Fe]}\geq0.95$). The number of stars in common exceeds 2000 for [Fe/H] and [C/Fe], but is only $\sim$300 for [$\alpha$/Fe]. This smaller sample reflects the fact that [$\alpha$/Fe] is the most difficult of the three parameters to recover from medium-band photometry, and therefore its photometric quality flags are generally lower. 
This is because the absorption lines produced by the $\alpha$-elements are weaker and less prominent than those of iron or carbon, making them more difficult to capture with medium-band filters.
As noted above, stars in the MC regions were not included in the training sample; thus, the APOGEE comparison provides an independent validation, although we caution that applying relations trained on MW stars to MC populations may introduce systematics due to their distinct star-formation and chemical-evolution histories.

As shown in Figure~\ref{fig:comp_mc}, the photometric [Fe/H] exhibits a moderate offset of $+0.17$\,dex relative to APOGEE (APOGEE minus photometry), with a scatter of $\sim$0.35\,dex. The photometric [C/Fe] is largely consistent with APOGEE, showing only a minor offset and a scatter of 0.33\,dex. For [$\alpha$/Fe], a moderate offset of $-0.09$\,dex is found, especially at [$\alpha$/Fe]~$< +0.1$, with a scatter of 0.07\,dex. Overall, the performance in the MC regions is somewhat worse than in the main survey area. This degradation primarily arises from the higher reddening toward the MCs, which can currently only be corrected using a two-dimensional map, and is further compounded by the slightly lower photometric precision in these crowded regions. In addition, the chemical patterns of stars in the MCs are partly affected by the limited coverage of the training sets (Figures~\ref{fig:training_feh} and \ref{fig:training_ca}), since these stars were optimized for the MW, rather than for the two dwarf galaxies. Nevertheless, the low metallicity and $\alpha$-abundance trends identified by the photometric technique remain consistent with expectations.

\subsection{Caveats and Limitations}
\begin{figure*}
\begin{center}
\includegraphics[scale=0.45,angle=0]{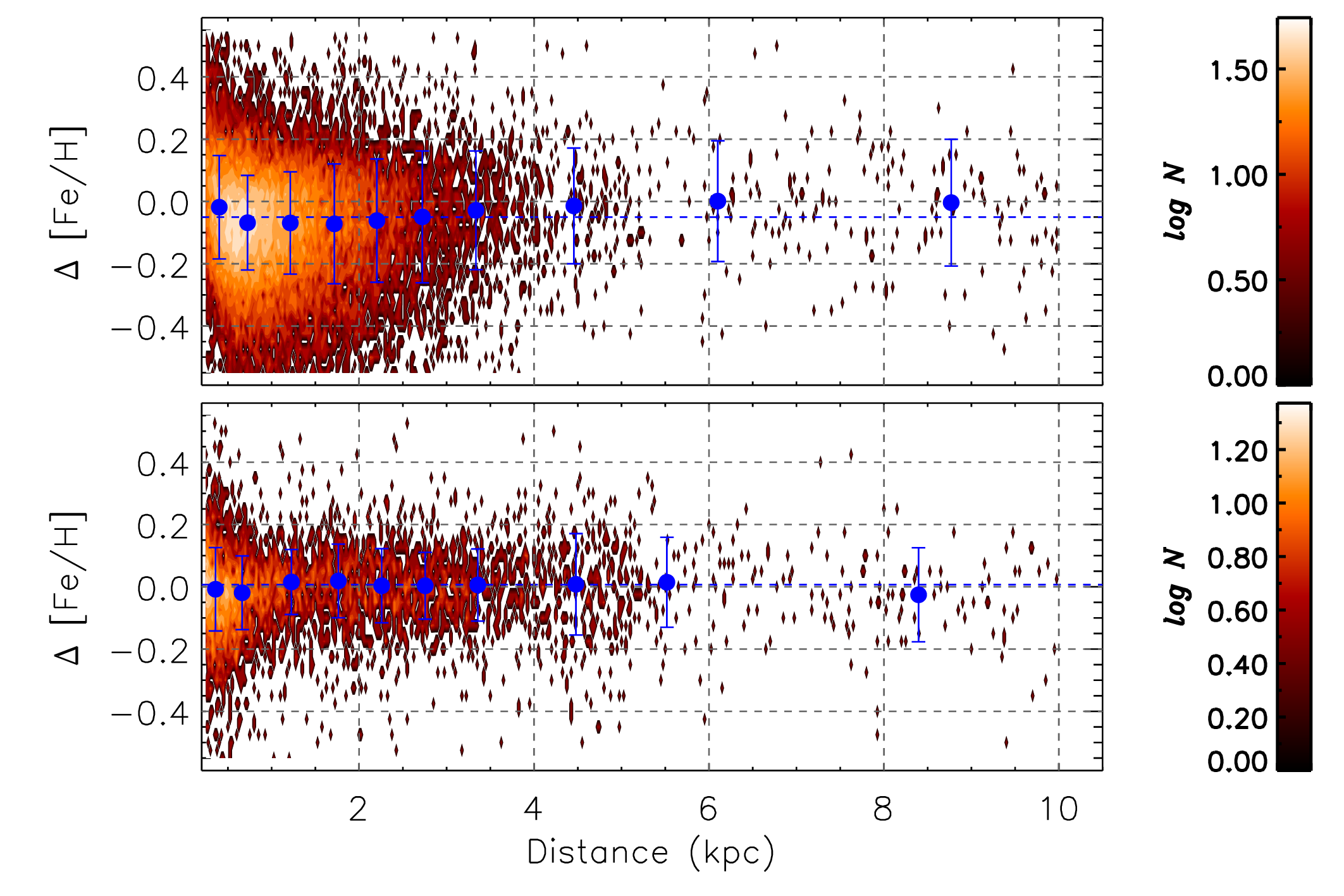}
\caption{Metallicity differences between photometric and spectroscopic estimates as a function of stellar distance.
The upper panel shows common stars between S-PLUS and LAMOST DR11 selected with \texttt{flg${\rm [Fe/H]} > 0.9$}, SNR $>20$, and $E(B-V)>0.03$, while the lower panel presents the comparison with APOGEE DR17 selected with \texttt{flg${\rm [Fe/H]} > 0.9$}, SNR $>50$, and $E(B-V)>0.03$.
Blue points with error bars denote the median and dispersion of the differences in each distance bin.
The blue dashed lines indicate the overall offset between photometric and spectroscopic metallicities.
The color bar on the right represents the stellar number density.}
\label{fig:comp_dist_dep}
\end{center}
\end{figure*}

We adopted the two-dimensional SFD reddening map \citep{SFD98} for extinction correction, as 92\% of the sample stars lie at Galactic latitudes $|b|>10^\circ$. 
While the overall extinction is modest at such latitudes, reddening is intrinsically distance dependent and therefore cannot be fully captured by a two-dimensional map. 
This limitation may introduce distance-related systematics into the photometric metallicities.
To quantify this effect, we cross-match our final catalog with LAMOST DR11 and APOGEE DR17 \citep{2022ApJS..259...35A}. 
Applying quality cuts of \texttt{flg$_{\rm [Fe/H]} > 0.9$} for S-PLUS, S/N $>20$ for LAMOST, and S/N $>50$ for APOGEE yields more than 26,000 (LAMOST) and nearly 6,000 (APOGEE) common stars with moderate extinction ($E(B-V)\ge0.03$). 
Figure~\ref{fig:comp_dist_dep} presents the difference between photometric and spectroscopic metallicities as a function of distance. 
No significant distance-dependent bias is observed, and any systematic offset remains within $\sim0.05$ dex, although the scatter increases moderately towards larger distances.
Future improvements will include the use of three-dimensional extinction maps and spatially varying extinction laws, which are expected to reduce both systematic and random uncertainties. 
This is particularly relevant for the LMC/SMC regions, where strong spatial variations in the extinction law have recently been reported \citep[e.g.,][]{2025Sci...387.1209Z}. 
Such refinements may significantly improve the accuracy of stellar parameter measurements in the Magellanic Clouds.

\section{Effective Temperatures, Distances, Ages, Surface Gravities and The S-PLUS stellar parameter sample}
Using data from S-PLUS DR4 and $Gaia$ EDR3, we estimate photometric metallicities, carbon-to-iron abundance ratios, and alpha-to-iron abundance ratios for about 5.4 million stars from the main survey and 0.7 million stars from the MCs, requiring quality flags of {\tt flg$_x > 0.6$}, where $x$ refers to [Fe/H], [C/Fe], or [$\alpha$/Fe]. For all main-survey stars, we further derive photometric effective temperatures $T_{\rm eff}$ using the de-reddened color $(G_{\rm BP} - G_{\rm RP})_0$ and photometric [Fe/H], adopting the metallicity-dependent $T_{\rm eff}$--color relations introduced in Paper~I. Distances are assigned following the same methodology as in Papers~I--III. For distant giants lacking reliable \textit{Gaia} parallaxes, we cross-match with the SMSS DR4 to obtain $gri$ photometry, from which absolute magnitudes and distances are derived using the empirical, metallicity-dependent color--magnitude relations constructed in Paper~I. Stellar ages are then estimated via a Bayesian approach, again as in Paper~I-III, combining constraints from $(G_{\rm BP} - G_{\rm RP})_0$, \textit{Gaia} $G$-band absolute magnitudes, and photometric metallicities.
We note that uncertainties in metallicity and extinction propagate non-linearly into age estimates, and are particularly significant for stars older than $\sim 10$ Gyr.
Surface gravities $\log g$ are also determined by isochrone fitting. In total, nearly 3.4 million stars are assigned ages, surface gravities, and luminosity classifications.

The typical uncertainty in the derived effective temperature is within 100 K when compared with spectroscopic measurements. Based on over 46,000 stars in common, the effective temperatures estimated in this work show excellent agreement with those from LAMOST, exhibiting a small offset of about 16 K (LAMOST minus this work) and a scatter of only 71 K (see Figure~\ref{fig:comp_tg}).
Similarly, the isochrone-based surface gravity values are consistent with those from LAMOST (Figure~\ref{fig:comp_tg}), showing a minor offset of $-0.03$~dex (LAMOST minus this work) and a scatter of 0.12\,dex. The isochrone-based ages are generally reliable for subgiant and turn-off stars, with a typical uncertainty of about 20\%. The distances of distant red giant stars derived from metallicity-dependent color–magnitude relations have typical uncertainties of 16\%. All of these results are comparable to those reported in Papers I–III of this series.

The spatial coverage and magnitude distributions of stars from the main survey and the MCs are presented in Figure~\ref{fig:distri}. Figure~\ref{fig:distri_param} further shows the distributions of [$\alpha$/Fe] versus [Fe/H] and [C/Fe] versus [Fe/H] for our sample. As expected, metal-rich thin-disk stars have [$\alpha$/Fe] values close to Solar, while metal-poor thick-disk and halo stars are $\alpha$-enhanced. Consistent with previous studies, carbon-enhanced stars ([C/Fe] $> +0.7$) are predominantly found in the metal-poor regime ([Fe/H] $< -1.0$), and are therefore classified as CEMP stars. In agreement with earlier results, the fraction of CEMP stars increases strongly toward decreasing [Fe/H]. Unlike MW stars, MC member stars exhibit distinct abundance trends. 
In the [$\alpha$/Fe] versus [Fe/H] plane, the $\alpha$-knee appears at [Fe/H] $\sim -2.4$, significantly lower than that of the Sgr dwarf galaxy \citep[e.g.,][]{2014MNRAS.443..658D}, despite Sgr being less massive than the MCs. 
This suggests a slower chemical-enrichment history in the MCs, consistent with predictions from 
chemical-evolution models \citep[e.g.,][]{2020ApJ...895..138K}.
In the [C/Fe] versus [Fe/H] plane, [C/Fe] decreases toward lower metallicity, opposite to the behavior of MW stars.
Most recently, \citet{2024NatAs...8..637C} identified ten stars in the LMC with [Fe/H] between $-4.2$ and $-2.5$, none of which are CEMP stars. In addition, \citet{2025ApJ...989L..18L} reported an ultra metal-poor (UMP; [Fe/H]~$<-4.0$) star possibly originating from the LMC with [Fe/H]~$=-4.82$, which is also carbon poor.
We note that several CEMP candidates have been found in the low-metallicity regime, warranting further spectroscopic follow-up to confirm their nature and provide crucial insights into the chemical evolution of dwarf galaxies.
These main trends were also reported by \citet{2020ApJ...895...88N} and \citet{2021ApJ...923..172H} based on abundance measurements from APOGEE. For comparison, we over-plot results for stars in common between APOGEE and the S-PLUS members; the trends are consistent with those derived from our photometric estimates.

Table 2 summarizes the contents of our final parameter sample. Astrometric information (i.e., parallaxes, proper motions, and their uncertainties) is taken from Gaia EDR3 \citep{GEDR3}, while the available radial velocities are compiled from various sources. The complete sample will be made publicly available at \url{https://zenodo.org/records/17411372}.

\section{Summary}

In this paper, we determine stellar parameters (effective temperature, surface gravity, [Fe/H], and age), along with key elemental-abundance ratios ([C/Fe] and [$\alpha$/Fe]), for more than 6 million stars. 
The sample includes 4.9 million dwarf stars and 0.5 million giant stars from the main survey, as well as 0.7 million red giant stars from the MCs. These estimates are based on 13 colors derived from a combination of narrow- and medium-band filter photometry from S-PLUS DR4 and ultra broad-band photometry from $Gaia$ EDR3. Using S-PLUS narrow/medium-band filters $J0395$ (tracing [Fe/H]) and $J0430$ (tracing [C/Fe]), we are able to disentangle metallicity from carbonicity, a capability already demonstrated in J-PLUS. This capability is particularly important for very metal-poor stars, where large fractions of carbon-enhanced objects have previously confounded photometric metallicity estimates in other surveys (e.g., SAGES, SMSS, and the Pristine survey). Typical uncertainties are 0.10–0.20\,dex for [Fe/H] and [C/Fe], and 
0.05\,dex for [$\alpha$/Fe], over much of the metallicity range. Our photometric [Fe/H] estimates remain reliable down to [Fe/H] $\sim -4.0$, with precisions of 0.40\,dex for dwarfs and 0.25\,dex for giants, and do not exhibit significant offsets. Together with earlier J-PLUS results from this series, the present sample enables robust studies of the metallicity distribution function and the frequency of CEMP stars across different disk and halo populations, based on a large and relatively unbiased stellar catalog. 

Applications of these samples have already proven successful, including the identification of dynamically and chemo-dynamically tagged groups and their connections to known substructures (e.g., \citealt{Shank2023}; \citealt{Zepeda2023}; \citealt{Cabrera2024}), as well as the discovery of candidate very and extremely metal-poor stars in the Galactic disk (e.g., \citealt{Hong2024}, and references therein). 
Our catalogs will also provide valuable input for selecting targets of particular interest for future medium- and high-resolution spectroscopic follow-up studies.

\section*{Acknowledgments}

The Guoshoujing Telescope (the Large Sky Area Multi-Object Fiber Spectroscopic Telescope, LAMOST) is a National Major Scientific Project built by the Chinese Academy of Sciences. Funding for the project has been provided by the National Development and Reform Commission. LAMOST is operated and managed by the National Astronomical Observatories, Chinese Academy of Sciences.

Y.H. acknowledges the support from the National Science Foundation of China (NSFC grant No. 12422303), the Fundamental Research Funds for the Central Universities (grant Nos. 118900M122, E5EQ3301X2, and E4EQ3301X2), and the National Key R\&D Programme of China (grant No. 2023YFA1608303). T.C.B. acknowledges partial support from grants PHY 14-30152; Physics Frontier Center/JINA Center for the Evolution of the Elements (JINA-CEE), and OISE-1927130; The International Research Network for Nuclear Astrophysics (IReNA), awarded by the US National Science Foundation, and DE-SC0023128; the Center for Nuclear Astrophysics Across Messengers (CeNAM), awarded by the U.S. Department of Energy, Office of Science, Office of Nuclear Physics. 
H.W.Z. acknowledges the support from the National Key R\&D Programme of China (grant No.
2024YFA1611903).
BD acknowledges support from ANID Basal project FB210003.
AAC acknowledges financial support from the Severo Ochoa grant CEX2021-001131-S funded by MCIN/AEI/10.13039/501100011033 and the project PID2023-153123NB-I00 funded by MCIN/AEI.
Y.S.L. acknowledges support from the National Research Foundation (NRF) of Korea grant funded by the Ministry of Science and ICT (RS-2024-00333766).
M.S.C. acknowledges support from São Paulo Research Foundation (FAPESP) grant 2025/12629-8.
P.K.H. gratefully acknowledges the Fundação de Amparo à Pesquisa do Estado de São Paulo (FAPESP) for the support grant 2023/14272-4.

\appendix
\setcounter{table}{0}   
\setcounter{figure}{0}
\renewcommand{\thetable}{A\arabic{table}}
\renewcommand{\thefigure}{A\arabic{figure}}
\section{Selecting MC member stars}

\begin{figure*}
\begin{center}
\includegraphics[scale=0.20,angle=0]{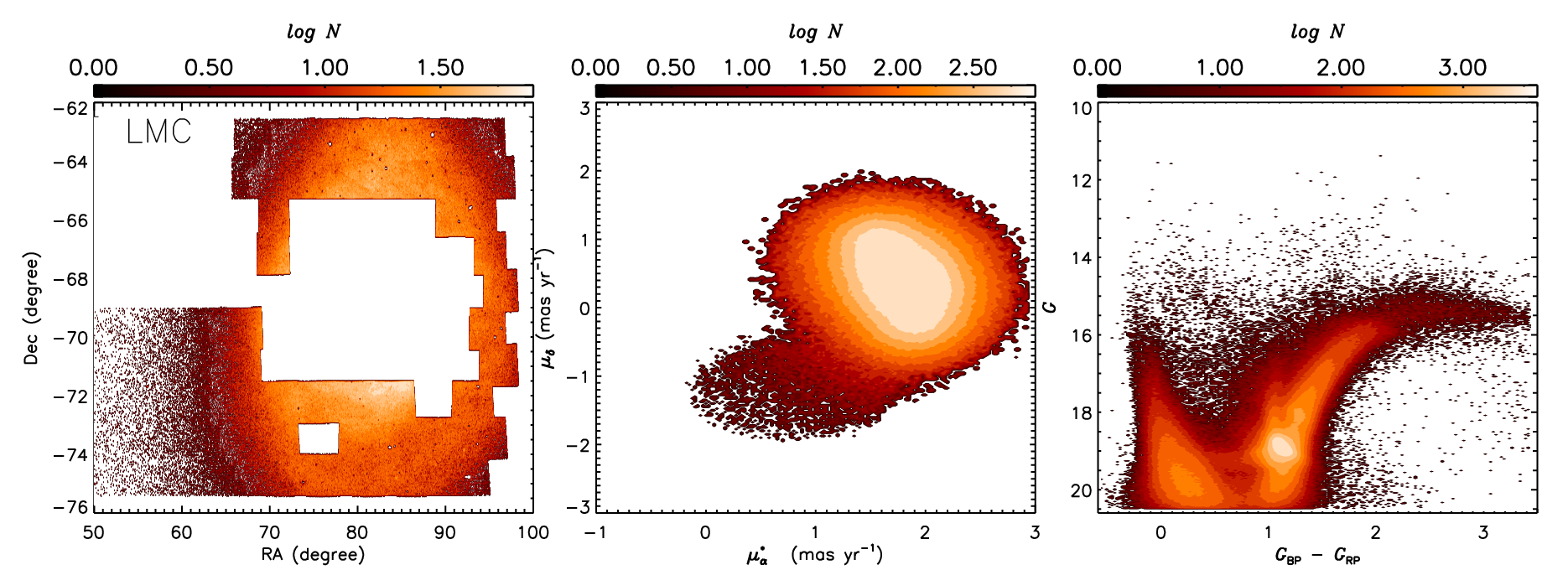}
\includegraphics[scale=0.20,angle=0]{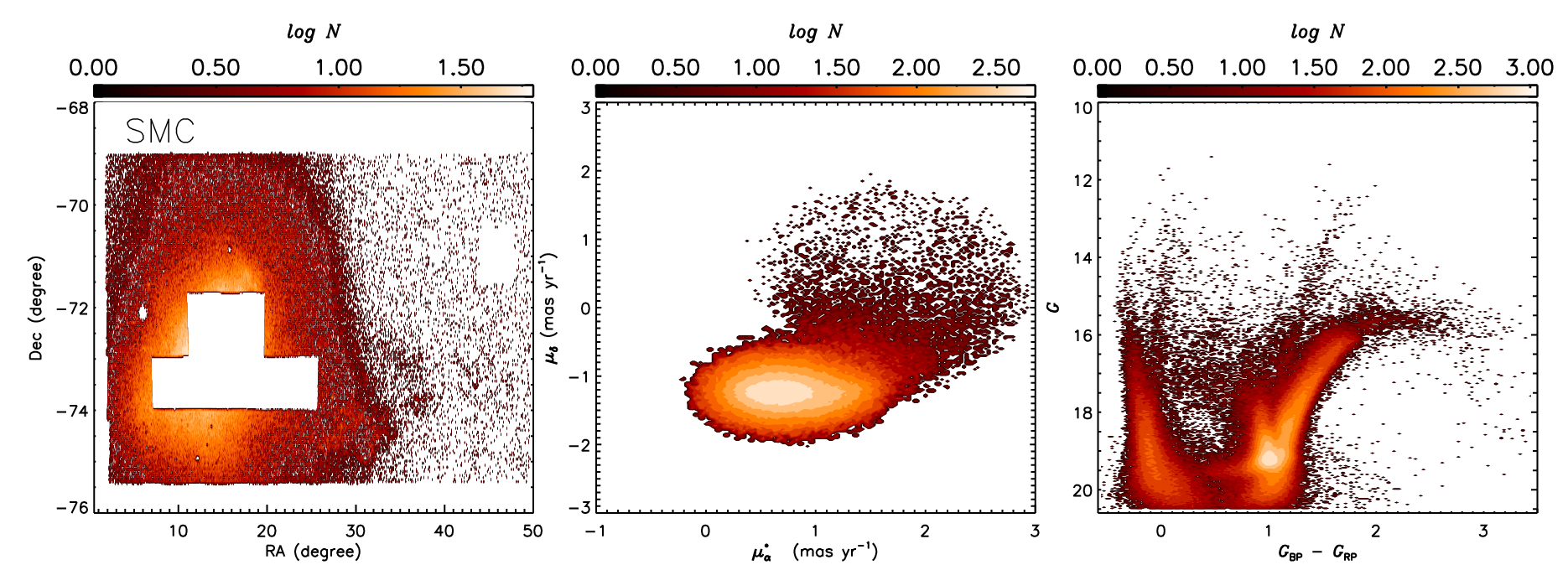}
\caption{Distributions of the selected MC member stars based on aperture photometry: spatial positions (left panels), proper motions (middle panels), and color–magnitude diagrams (right panels). The upper panels show the LMC stars, while the lower panels correspond to the SMC. A color bar indicating the number density is shown at the top of each panel.}
\label{fig:mc_aper}

\end{center}
\end{figure*}

\begin{figure*}
\begin{center}
\includegraphics[scale=0.20,angle=0]{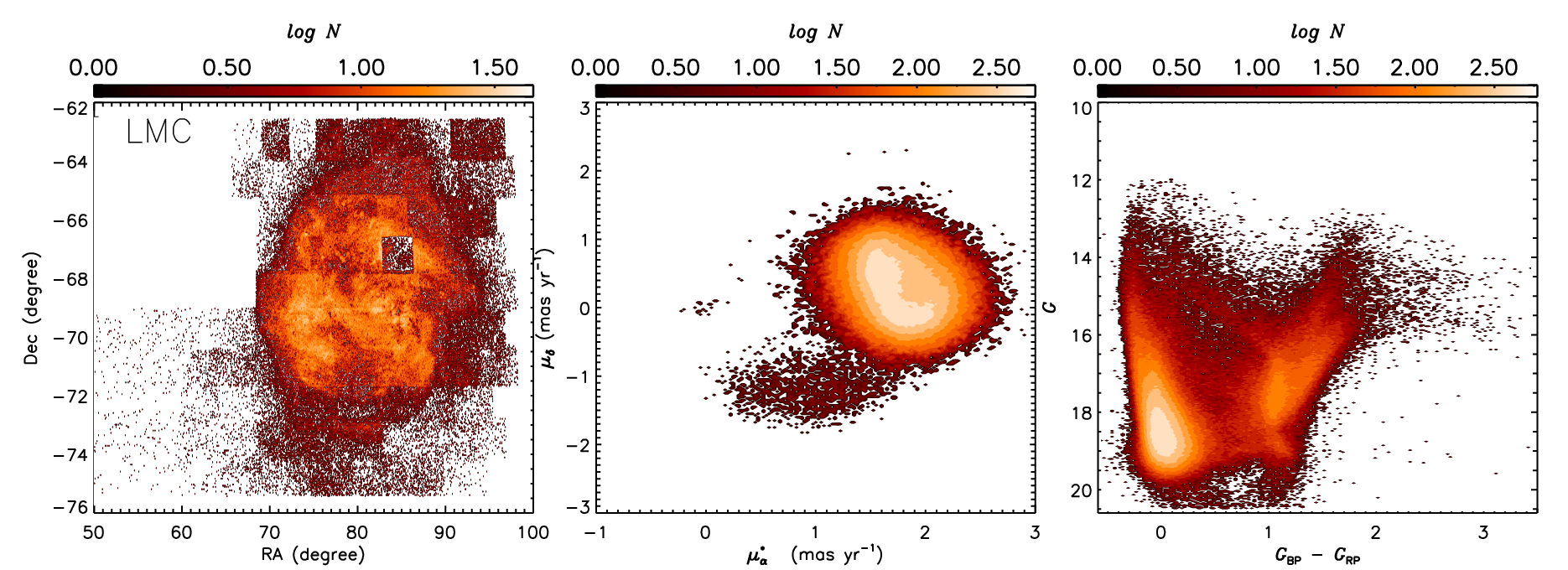}
\includegraphics[scale=0.20,angle=0]{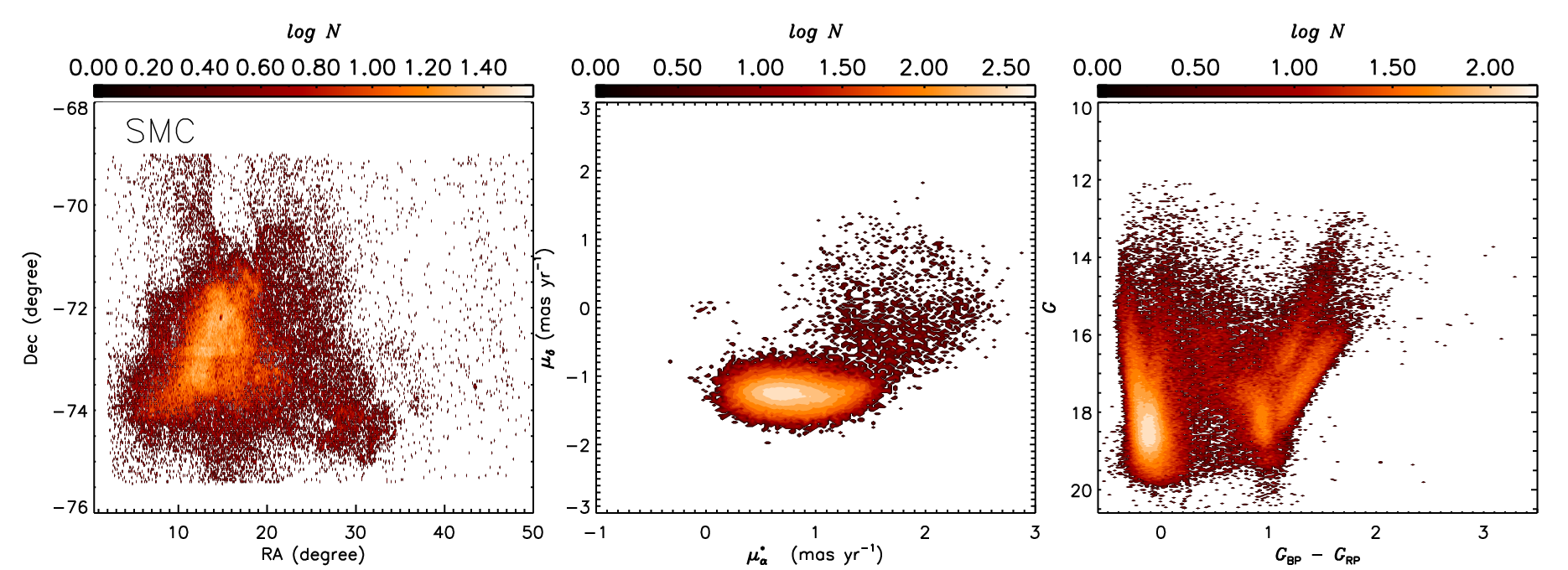}
\caption{Similar to Figure~\ref{fig:mc_aper}, but for member stars selected using PSF photometry. A color bar indicating the number-density is shown at the top of each panel.}
\label{fig:mc_psf}
\end{center}
\end{figure*}

Using $Gaia$ parallaxes as priors, we take advantage of the fact that stars with large parallaxes (i.e., at close distances) are mostly foreground field stars along the line of sight to the Magellanic Clouds (MCs).  
We therefore divide all stars in the MC directions into different parallax (distance) bins and examine their distributions in the two-dimensional proper-motion diagram. MC member stars appear as two distinct overdensities, one centered at ($\mu_{\alpha}^*$, $\mu_{\delta}$) $\approx$ ($1.9$, $0.2$) mas yr$^{-1}$ corresponding to the LMC, and the other at ($0.8$, $-1.2$) mas yr$^{-1}$ corresponding to the SMC.  
To account for varying levels of contamination, we adopt different proper-motion cuts for different parallax bins to exclude foreground stars.  
In total, nearly 2 million MC member stars are identified using both aperture photometry (mainly covering the outer regions of the MCs; see Figure~\ref{fig:mc_aper}) and PSF photometry (including the crowded inner regions; see Figure~\ref{fig:mc_psf}).  
The color-magnitude diagram shows that these member stars are predominantly located in the upper part of the diagram, i.e., hot blue main-sequence stars and red giant branch stars.  
At present, our stellar-parameter estimation technique is most effective for red giant branch stars.

\begin{figure*}
\begin{center}
\includegraphics[scale=0.35,angle=0]{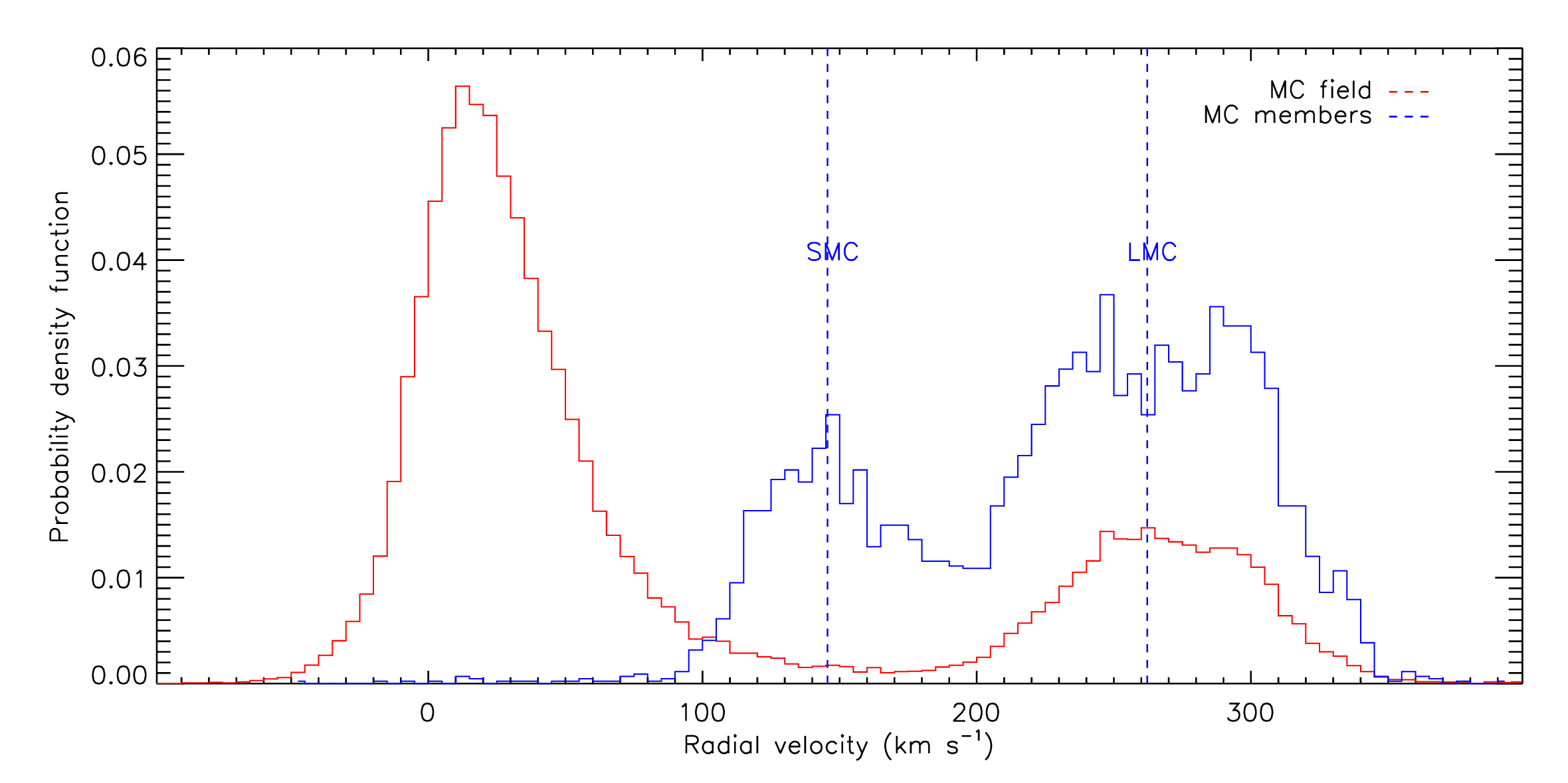}
\caption{Radial velocity distributions of all 54,905 MC field stars in APOGEE DR17 (SNR $> 20$; red line) and of proper-motion-selected S-PLUS member stars with radial velocity measurements from APOGEE DR17 (SNR $> 20$; blue line).  
The two blue-dashed lines indicate the systemic radial velocities of $262.2$\,km\,s$^{-1}$ for the LMC \citep{2002AJ....124.2639V} and $145.6$\,km\,s$^{-1}$ for the SMC \citep{2006AJ....131.2514H}.} 
\label{fig:mc_mem}
\end{center} 
\end{figure*}

To assess the purity of the selected MC member stars, Figure~\ref{fig:mc_mem} shows the radial velocity distribution of stars in common between S-PLUS and APOGEE DR17 \citep{2022ApJS..259...35A}.  
The distribution clearly exhibits two distinct peaks, in good agreement with the systemic radial velocities of the LMC and SMC reported in the literature \citep{2002AJ....124.2639V, 2006AJ....131.2514H}.  
For comparison, we also show the radial velocity distribution of all stars in the MC directions targeted by the APOGEE survey.  
As is evident, a large fraction of foreground stars, with radial velocities between $-50$ and 100\,km\,s$^{-1}$, 
dominate the distribution.  These results confirm the success of our proper-motion-based MC member star selection.

\clearpage
\vfill\eject
\bibliography{sppara_calib}{}
\bibliographystyle{aasjournal}
\end{document}